%% file: FiniteField.tex
\documentclass[11pt,draftclsnofoot,peerreviewca,letterpaper,onecolumn]{IEEEtran}
\usepackage{cite}

\usepackage{graphicx,color,epsfig,rotating,subfigure}
\usepackage{amsfonts,amsmath,amssymb}
\usepackage{algorithm,algorithmic}
\usepackage{subfigure}

\input{macros}

\newtheorem{definition}{Definition}
\newtheorem{theorem}{Theorem}
\newtheorem{lemma}{Lemma}
\newtheorem{corollary}{Corollary}

\newtheorem{remark}{Remark}

\newtheorem{example}{Example}

\begin{document}

\sloppy

\title{On Interference Networks over Finite Fields}

\author{\authorblockN{Song-Nam~Hong,~\IEEEmembership{Student Member,~IEEE,}
        and~Giuseppe~Caire,~\IEEEmembership{Fellow,~IEEE}}
\authorblockA{Department of Electrical Engineering, University of Southern California, Los Angeles, CA, USA}
\authorblockA{(e-mail: \{songnamh, caire\}$@$usc.edu)}}%


\maketitle

\begin{abstract}
We present a framework to study linear deterministic interference networks over finite fields. Unlike the popular linear deterministic models
introduced to study Gaussian networks, we consider networks where the channel coefficients are general scalars over some extension field
$\FF_{p^m}$ (scalar $m$-th extension-field models), $m \times m$ diagonal matrices over 
$\FF_p$ ($m$-symbol extension ground-field models), and $m \times m$ general non-singular matrices (MIMO ground field models). 
We use the companion matrix representation of the extension field to convert $m$-th extension scalar models into MIMO ground-field 
models where the channel matrices have special algebraic structure. For such models, we consider the $2 \times 2 \times 2$ topology (two-hops two-flow)
and the 3-user interference network topology. We derive achievability results and feasibility conditions for certain
schemes based on the Precoding-Based Network Alignment (PBNA) approach, where intermediate nodes use random linear network coding
(i.e., propagate random linear combinations of their incoming messages) and non-trivial precoding/decoding is performed only at the network edges,
at the sources and destinations. Furthermore, we apply this approach to the scalar $2\times 2\times 2$ complex Gaussian IC with fixed channel coefficients, 
and show two competitive schemes outperforming other known approaches at any SNR, where we combine 
finite-field linear precoding/decoding with lattice coding and the Compute and Forward approach at the signal level. 
As a side result, we also show significant advantages of vector linear network coding both in terms of feasibility probability (with random coding coefficients)
and in terms of coding latency, with respect to standard scalar linear network coding, in PBNA schemes.
\end{abstract}


\clearpage

\section{Introduction}  \label{sec:intro}

Recently, significant progress has been made towards the understanding of the theoretical limits of wired and wireless
communication networks. Most remarkable advances have been accomplished in two settings:
1) multiple multicasts relay networks, i.e., multihop networks where {\em every destination desires all messages};
2) multiple unicasts over a single hop, where each destination requires a message for its intended source.
Other mixed scenarios have also been considered in a less systematic manner, as for example the X channel (where each source has a message
for each destination) \cite{Maddah-Ali,Jafar-X,Jafar-X1}, or index coding, i.e., a deterministic broadcast channel where each destination is characterized by a set of desired messages and
a set of side-information messages \cite{Birk}.

The capacity of multiple multicast networks is exactly known for wired networks \cite{Ahlswede} and is approximately known
within a constant gap for wireless networks \cite{Avestimehr}.  Also, for multiple flows over a single hop, capacity approximations
were obtained in the form of degrees of freedom (DoF), generalized DoF (GDoF), and $O(1)$-gap bounds \cite{Cadambe08,Gou09,Jafar10}.
In particular, the concept of {\em interference alignment} (IA) was proposed by Cadambe and Jafar \cite{Cadambe08} to achieve
the optimal sum-DoF of the $K$-user Gaussian Interference Channel (IC) with time-varying coefficients (infinite channel diversity), equal to $K/2$.
Also in the assumption of infinite channel diversity, {\em ergodic} IA \cite{Nazer-Erg} was shown to achieve the {\em ergodic} capacity of 
time-varying $K$-user  Gaussian ICs within a bounded gap, under certain symmetry conditions on  the channel coefficients statistics. 
In the assumption of infinite channel resolution, a construction known as ``real IA'' \cite{Motahari}, where alignment is achieved by 
coding over rationally independent basis, 
was shown to achieve the optimal $K/2$ sum-DoF of the $K$-user Gaussian IC with {\em fixed} channel coefficients, 
with probability 1 over the channel coefficients drawn from a continuous distribution. 
A comprehensive survey of IA is provided in the monograph \cite{Jafar-Mo}.

Although the case of multiple flows over multiple hops remain widely unsolved in general, several important advances have been made recently. 
For the case of two flows, the two-hop network given by the concatenation of two two-user interference channels (referred to as the $2\times 2\times 2$ IC) 
has received much attention, being one of the fundamental building blocks to characterize the DoFs of general two flows networks \cite{Shomorony}.
The optimal sum-DoF of the $2\times 2\times 2$ Gaussian IC was obtained in \cite{Gou} using a technique known as {\em aligned interference neutralization},
which appropriately combines interference alignment and interference neutralization \cite{Mohajer}.  
Also, a recent extension to the $K \times K \times K$ Gaussian IC, achieving the optimal DoF equal to $K$ was
obtained in \cite{Shomorony1}, using a technique referred to as {\em aligned network diagonalization} (AND).


In this paper we focus on linear deterministic finite-field interference networks where channel coefficients and the input-output symbols along to
the same finite field. These models are different from the ``classical'' linear deterministic models introduced in  \cite{Avestimehr}, where
the input/output symbols are vectors over the binary field, and the channel coefficients are restricted to be ``shift'' matrices, expressing the attenuation/amplification
of the signal levels.  In general, finite-field models capture the key features of broadcast and (linear) superposition in networks, while
allow for a simple algebraic framework.  A few motivations for studying such networks are given in the following:
1) In practical wireless communication systems, the main bottleneck of a digital receiver is the Analog-to-Digital Conversion (ADC), which is power-hungry,
and does not scale with Moore's law. Rather, the number of quantization bits per second at the output of an ADC is nearly a constant that depends
on the power consumption \cite{Walden}. Thus, it makes sense to consider the ADC as part of channel, such that the channel output is intrinsically discrete
with number of bits per sample inversely proportional to the signal bandwidth, for a fixed receiver power consumption.
This means that it is impossible to achieve arbitrary resolution in the signal levels. In contrast, using appropriate lattice coding it is possible to
represent the channel as a linear finite-field model with discrete noise, as shown by the authors in \cite{Hong-DAS}. In this case, focusing to
the noiseless case may be seen as a high-SNR but finite resolution abstraction of the underlying network;
2) the general compute-and-forward (CoF) framework can be used to convert a Gaussian interference network into a
finite-field linear deterministic network  \cite{Hong-DAS,Nazer,Hong-J}. CoF makes use of lattice codes such that each intermediate node (relay)
can reliably decode a linear combination with integer coefficients of the interfering lattice codewords.
Thanks to the fact that lattices are modules over the ring of integers, such linear combinations translate directly into linear combinations of the information
messages defined over a suitable finite-field. By forwarding such linear combinations, the overall end-to-end network
``transfer function" between sources and destinations can be described by a system of deterministic linear equations over finite-field, where the noise in the network has been eliminated by lattice decoding at each intermediate node.  Then, each destination can solve such equations to obtain
its own desired messages, as long as there exists a full-rank sub-system of equations involving the desired messages.
3) in the case of wired (deterministic) network with linear network coding at the intermediate nodes,
an arbitrary network can be turned into a finite-field one-hop interference network
where all the encoding/decoding operations are restricted to be performed at the sources and destinations.
This approach is known as {\em precoding based network alignment} (PBNA).
For example, as shown in \cite{Das,Meng}, three unicast flows over wired networks can be converted into a 3-user finite-field IC with
diagonal channel matrices if all relays perform scalar linear network coding operations. 4) As noticed in \cite{Krishnamurthy}, the study of finite-field networks
can provide insights to solve important open problems in wireless networks such as finite SNR and finite channel diversity scenarios, going
beyond DoF results. For example, we observe that a single deterministic link over $\FF_{p^m}$ has capacity $m\log{p}$,
that is similar to have $m$ parallel channels of capacity $\log{p}$ each.  This establishes an intuitive
correspondence between $m$ (the field extension order) and ``channel diversity" (number of parallel channels) and
between $p$ (the ground field size) and SNR.

Interestingly, interference management schemes such as IA and AND,
originally developed for Gaussian interference networks, are not straightforwardly applicable to the finite-field case. This is because
in the Gaussian case the feasibility conditions of such schemes are satisfied with arbitrarily high probability
provided that the channel coefficients are independently drawn from some continuous distribution  \cite{Cadambe08,Gou,Shomorony1,Motahari}.
In the finite-field counterparts, however, signals and channel coefficients belong
to a finite-field and hence the feasibility conditions must be carefully considered. Further, the fraction of channel coefficients for which such schemes are feasible
can be a non-trivial function of $p$ and $m$. In this paper, our main goal is to formulate the feasibility conditions of interference management
schemes by taking the algebraic structures of channel matrices into account and derive the probability that feasibility conditions hold
(in short, {\em feasibility probability}) as function of $p$ and $m$.

\subsection{Contributions}

We present a novel framework to convert a $K$-user linear deterministic multiple access channel (MAC) over $\FF_{p^m}$ with single input/output into
a $K$-user MAC over ground-field $\FF_{p}$ with $m$ multiple inputs/outputs (MIMO).
In a work developed independently and in parallel to ours, Krishnamurthy and Jafar \cite{Krishnamurthy}
present the same idea and show that the resulting $m \times m$ channel matrices have additional structure such as commutativity 
with respect to matrix multiplication.\footnote{We would like to mention that \cite{Krishnamurthy}
focuses on the 2-user X channel and the 3-user IC, while here we focus on the $2\times 2 \times 2$ IC as well as on the 3-user IC.
About the concurrency and independence of \cite{Krishnamurthy} and our work, it is interesting to notice that
a pre-print of \cite{Krishnamurthy} appeared on ArXiv on April 29, 2013, while a conference version of the present paper was submitted to the
IEEE Information Theory Workshop on April 20, 2013.}
 In this paper, we further show that the $m\times m$ matrices or the transformed MIMO channel
are represented by the powers of the {\em companion matrix} of a primitive polynomial that generates the extension-field $\FF_{p^m}$. 
Thus, commutativity follows immediately as a consequence of the field multiplicative matrix representation.  
Our framework allows to develop coding schemes for MIMO
channels as done in symbol-extension (i.e., coding over multiple time slots) for time-varying 
channels.

We begin with the $2\times 2\times 2$ scalar IC over $\FF_{p^m}$ and show that the sum-rate of $(2m-1)\log{p}$ is achievable by applying
AND to a transformed $2\times 2\times 2$ MIMO IC, under certain feasibility conditions on the channel coefficients.
Also, we prove that these conditions hold with probability 1 if the channel coefficients are uniformly and independently drawn from the non-zero
elements of $\FF_{p^m}$ and either $m$ or $p$ goes to infinity.  Focusing on a $2\times 2\times 2$ MIMO IC over $\FF_{p}$, we show that symbol-extension
(i.e., coding over multiple time slots) is required in order to employ AND, differently from the Gaussian counterpart.
This is because complex-field is algebraically closed while finite-fields are not.
By comparing the feasibility probabilities of field-extension, symbol-extension, and MIMO, we observe that only field-extension
can guarantee feasibility with probability 1 for any fixed $p$ as long as $m$ large enough.
This observation motivates us to propose a {\em vector} random linear network coding (RLNC) for two-flow multihop wired networks.
In this scheme, coding matrices are chosen from the powers of a companion matrix of a primitive polynomial that generates an extension-field
$\FF_{p^m}$. That is, all relays use the same companion matrix and choose its powers uniformly and independently
from $\{0,1,\ldots,p^{m}-2\}$.
We show that vector RLNC with AND achieves the cut-set upper bound for certain network topologies.
In contrast, {\em scalar} RLNC with AND can achieve the same bound only when $p$ is sufficiently large (i.e., the capacity of a wired links is large enough).
This represents a very strong argument in favor of {\em vector} versus {\em scalar} RLNC. As an application of finite-field study, we consider a
$2\times 2\times 2$ Gaussian IC with {\em constant} channel coefficients and present two coding schemes referred to as
CoF-AND (Compute and Forward with Aligned Network Diagonalization) and
PCoF-CIA (Precoded Compute and Forward with Channel Integer Alignment).  They are constructed by appropriately combining the framework of
extension-field and CoF.   We show that CoF-AND outperforms time-sharing at lower range of SNRs and PCoF-CIA does so in a range of moderate SNR to high SNR,
with increasing gain as SNR increases.

For the 3-user MIMO IC over $\FF_{p}$, we show that eigenvector-based IA \cite{Cadambe08} is also feasible with an appropriate symbol extension.
Yet, we give a negative result for the {\em scalar} 3-user IC over $\FF_{p^m}$. Independently, the same negative result has been shown in \cite{Krishnamurthy}.
While in \cite{Krishnamurthy} the authors do not allow to find eigenvalues in an extension-field, i.e., they restrict to eigenvalues only in the ground-field, 
in this paper we provide a stronger result showing that eigenvector-based IA  is not feasible although we can always find eigenvalues in an extension-field
and coding/decoding can be performed over symbol-extension by going to the extension-field.
Finally, we show that using vector linear network coding based on the companion matrix approach
for PBNA schemes, as introduced in \cite{Meng,Das}, has some important advantages in terms of feasibility probability 
and  latency.

The paper is organized as follows.
In Section~\ref{sec:MAC}, we summarize some definitions on finite-field and present a novel framework to convert a SISO MAC over an extension-field
into a MIMO MAC over the corresponding ground-field.
Section~\ref{sec:SM} defines the relevant system models corresponding to different algebraic structures of the channel matrices.
In Section~\ref{sec:FE}, we formulate the feasibility conditions of AND for the $2\times 2\times 2$ IC over $\FF_{p^m}$ and derive the feasibility
probability as function of $p$ and $m$. Also, we prove that this probability approaches to 1 if either $m$ or $p$ goes to infinity.
Section~\ref{sec:MIMO} shows  that AND can be applied to a $2\times 2\times 2$ MIMO IC over $\FF_{p}$ with proper symbol-extension
and we also derive the corresponding feasibility probability.
In Section~\ref{sec:PBNA-2}, we propose a vector RLNC with AND for two-flow multihop wired networks and show that this scheme achieves
the cut-set upper bound for certain network topologies.
In Section~\ref{sec:GA}, as an application, we present CoF-AND and PCoF-CIA for a $2\times 2\times 2$ Gaussian IC with constant channel
coefficients. In Section~\ref{sec:3-User}, we show that an eigenvector-based IA scheme is feasible for the 3-user MIMO IC over $\FF_{p}$, provided that 
coding takes place on an appropriate splitting field extension of $\FF_p$, that translates in coding over a symbol extension over the original ground field. 
However, such scheme is not feasible for the 3-user {\em scalar} IC over $\FF_{p^m}$  although we allow symbol extensions of any order. 
Some concluding remarks are provided in Section~\ref{sec:con}.


\section{MIMO Representation of field-extension}\label{sec:MAC}

We first provide some basic notations and definitions that will be used in the sequel. Let $p$ be a prime and $\FF_{p^m}$ denote a finite-field of order $p^m$ with elements
\begin{equation}
\{0,1,\alpha,\alpha^2,\ldots,\alpha^{p^{m}-2}\}
\end{equation} where $\alpha$ is a {\em primitive element}. As usual, $\FF_{p^m}^{\star}$ denotes the multiplicative group of $\FF_{p^m}$, i.e., the set of non-zero elements of $\FF_{p^{m}}$.

\begin{definition}
Let $\beta \in \FF_{p^m}$. The {\em minimal polynomial} of $\beta$ with respect to $\FF_{p}$ is the lowest degree {\em monic} polynomial $\pi_{\beta}(x)$ with coefficients in $\FF_{p}$ such that $\pi_{\beta}(\beta) = 0$. Also, we let $\deg(\pi_{\beta}(x))$ denote the degree of polynomial $\pi_{\beta}(x)$. For a primitive element $\alpha$,  the minimal polynomial $\pi_{\alpha}(x)$ has degree $m$ and is called {\em primitive polynomial}.
\hfill $\lozenge$
\end{definition}

Given a primitive element $\alpha$ with primitive polynomial
\begin{equation}
\pi_{\alpha}(x) \eqdef a_{0} + a_{1}x + \ldots + a_{m-1}x^{m-1} + x^{m},
\end{equation}
the condition $\pi_{\alpha}(\alpha) = 0$ yields the recursive representation
\begin{equation}
\alpha^{m} = -a_{0}-a_{1}\alpha - \ldots - a_{m-1}\alpha^{m-1}.
\end{equation} Using this relation, we can put in one-to-one correspondence the field of polynomials over the ground field $\FF_{p}$ modulo $\pi_{\alpha}(x)$, denoted by $\FF_{p}[x]/\pi_{\alpha}(x)$ with the vector space of dimension $m$ over the ground field $\FF_{p}$, and with the extension field $\FF_{p^m}$.
That is, we can define the one-to-one mapping $\Phi : \FF_{p^m} \rightarrow \FF_p^m$ (called \textbf{Vector representation}), given by:
\begin{equation}
\Phi(\alpha^{\ell}) = [b_{0},b_{1},\ldots,b_{m-1}]^{\transp}\label{def:Phi}
\end{equation} where $\alpha^{\ell} = b_{0}+b_{1}\alpha+\ldots+b_{m-1}\alpha^{m-1}$.

\begin{definition} The {\em companion matrix} of a primitive polynomial $\pi_{\alpha}(x)=a_{0} + a_{1}x + \cdots + a_{m-1}x^{m-1} + x^{m}$ is the $m \times m$ matrix over $\FF_{p}$
\begin{equation*}
\Cm\eqdef\left[
    \begin{array}{ccccc}
      0 & 0 & \cdots& 0 & -a_{0} \\
      1 & 0 & \cdots &0 & -a_{1}\\
      0 & 1 & \cdots & 0 & -a_{2} \\
      \vdots & \vdots &\ddots &\vdots &\cdots\\
      0 & 0 & \cdots & 1 & -a_{m-1} \\
    \end{array}
  \right].
\end{equation*}
 \hfill $\lozenge$
\end{definition} Then, $\Cm$ is a generator of the cyclic group under matrix multiplication
\begin{equation}
\Cc^{\star}(p,m) \eqdef \{\Id = \Cm^{0},\Cm,\ldots,\Cm^{p^{m}-2}\}.
\end{equation} We know that the characteristic polynomial of $\Cm$, i.e., $\det(\Cm-\lambda\Id)$, is equal to the primitive
polynomial $\pi_{\alpha}(x)$.  Hence, we have that $\pi_{\alpha}(\Cm) = \mbox{0}$. This condition yields recursive representation:
\begin{equation}
\Cm^{m} = -a_{0}\Id-a_{1}\Cm-\cdots -a_{m-1}\Cm^{m-1}.
\end{equation}
Using this relation, we can show that $\Cc(p,m) = \Cc^{\star}(p,m) \cup \{\mbox{0}\}$ forms a commutative group under addition, such that
the set of matrices
\begin{equation}
\Cc(p,m) \eqdef \{\mbox{0} \eqdef  \Cm^{\infty},\Id=\Cm^{0},\Cm,\ldots,\Cm^{p^{m}-2}\}
\end{equation}
is a field under matrix addition and multiplication.
From \cite[Th. 6]{Mac}, all finite-fields of order $p^m$ are isomorphic.\footnote{Two fields $F,G$ are said to be {\em isomorphic} if there is a one-to-one mapping from $F$ onto $G$ which preserves addition and multiplication.}
Thus, $\Cc(p,m)$ and $\FF_{p^m}$ are isomorphic and we have the one-to-one mapping
$\Gamma : \FF_{p^m} \rightarrow \Cc(p,m)$ (called \textbf{Matrix representation}), defined by:
\begin{equation}
\Gamma(\alpha^{\ell}) = \Cc(p,m)  \Cm^{\ell}\label{def:Gamma}.
\end{equation}

\begin{example}
Take $\pi_{\alpha}(x) = x^3+x+1$ over $\FF_{2}$. Its companion matrix is given by
\begin{equation}
\Cm=\left[
      \begin{array}{ccc}
        0 & 0 & 1 \\
        1 & 0 & 1 \\
        0 & 1 & 0 \\
      \end{array}
    \right],
\end{equation}
The set $\Cc(p,m) = \{\mbox{0},\Id,\Cm,\ldots,\Cm^{6}\}$ form a field isomorphic to $\FF_8$. From the identity
$\Cm^3 + \Cm + \Id = 0$ we can calculate addition. For example,
\begin{eqnarray*}
\Cm^{4} + \Cm^{5} &=& (\Cm^2 + \Cm) + (\Cm^2 + \Cm + \Id)=\Id.
\end{eqnarray*}
Equivalently, we have:
\begin{eqnarray*}
\Cm^{4} + \Cm^{5} &=& \left[
                        \begin{array}{ccc}
                          0 & 1 & 1 \\
                          1 & 1 & 0 \\
                          1 & 1 & 1 \\
                        \end{array}
                      \right] + \left[
                        \begin{array}{ccc}
                          1 & 1 & 1 \\
                          1 & 0 & 0 \\
                          1 & 1 & 0 \\
                        \end{array}
                      \right] = \left[
                                  \begin{array}{ccc}
                                    1 & 0 & 0 \\
                                    0 & 1 & 0 \\
                                    0 & 0 & 1 \\
                                  \end{array}
                                \right].
\end{eqnarray*}\hfill $\lozenge$
\end{example}

\begin{figure}[ht]
\centerline{\includegraphics[width=16cm]{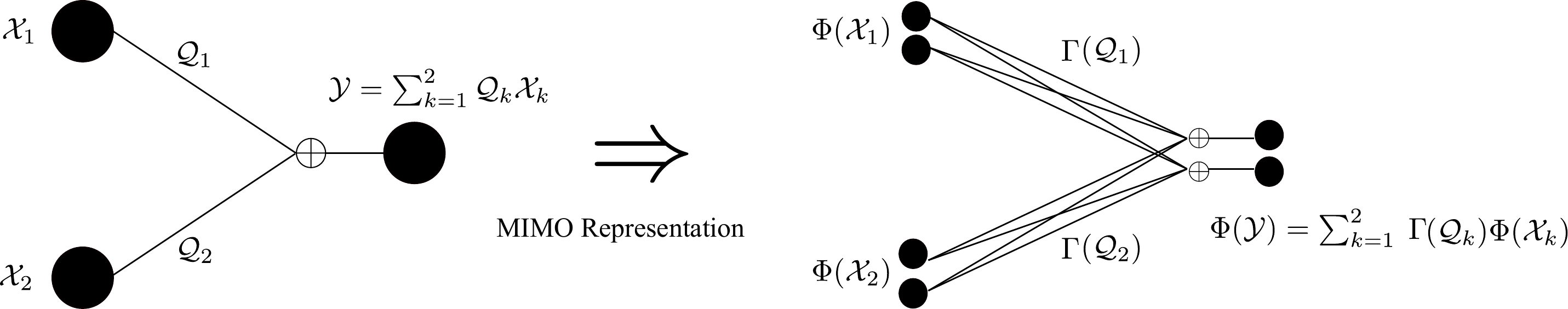}}
\caption{MIMO Transform from scalar MAC over $\FF_{p^2}$ into MIMO MAC over $\FF_{p}$.}
\label{mimoTR}
\end{figure}

The following result presents a framework to convert a SISO network over $\FF_{p^m}$ into a $m \times m$ MIMO network over ground field $\FF_{p}$ (see Fig.~\ref{mimoTR}):

\begin{lemma}\label{lem1}
Let  $\Yc = \sum_{k=1}^{K} \Qc_{k} \Xc_{k}$ denote a deterministic linear MAC over $\FF_{p^m}$, such that
both the inputs $\{\Xc_{k}\}$ and the channel coefficients $\{\Qc_k\}$ are elements of $\FF_{p^m}$.  Also, let
$\yv = \sum_{k=1}^{K}\Qm_{k}\xv_{k}$,  where $\xv_{k} = \Phi(\Xc_{k}) \in \FF_{p}^{m}$ and  $\Qm_{k} = \Gamma(\Qc_{k}) \in \FF_{p}^{m \times m}$,
for $k=1,\ldots,K$. Then, we have $\yv = \Phi(\Yc)$.\hfill\IEEEQED
\end{lemma}

\begin{example} Take $\pi_{\alpha}(x) = x^2 + x + 1$ over $\FF_{2}$.
The elements of $\FF_{4}$ are $\{0,1,\alpha,\alpha^2\}$. The corresponding  matrix representation using the companion matrix $\Cm$ is given by:
\begin{equation}
\begin{array}{cccc}
  \textbf{0} & \Cm & \Cm^2 & \Cm^3 = \Id \\
  \left[
         \begin{array}{cc}
           0 & 0 \\
           0 & 0 \\
         \end{array}
       \right] & \left[
         \begin{array}{cc}
           0 & 1 \\
           1 & 1 \\
         \end{array}
       \right] & \left[
         \begin{array}{cc}
           1 & 1 \\
           1 & 0 \\
         \end{array}
       \right] &  \left[
         \begin{array}{cc}
           1 & 0 \\
           0 & 1 \\
         \end{array}
       \right].
\end{array}
\end{equation}
The vector and matrix representations are given by
\begin{eqnarray*}
\Phi(\{0,1,\alpha,\alpha^2=1+\alpha\}) = \{[0,0]^{\transp}, [1,0]^{\transp}, [0,1]^{\transp}, [1,1]^{\transp}\},
\end{eqnarray*}
and by
\begin{eqnarray*}
&&\Gamma(\{0,1,\alpha,\alpha^2=1+\alpha\})=\left\{\left[
         \begin{array}{cc}
           0 & 0 \\
           0 & 0 \\
         \end{array}
       \right], \left[
         \begin{array}{cc}
           1 & 0 \\
           0 & 1 \\
         \end{array}
       \right], \left[
         \begin{array}{cc}
           0 & 1 \\
           1 & 1 \\
         \end{array}
       \right],\left[
         \begin{array}{cc}
           1 & 1 \\
           1 & 0 \\
         \end{array}
       \right]
   \right\},
\end{eqnarray*}
respectively.  Consider the two-user MAC over $\FF_{4}$ with
\begin{equation}
\Yc = \Qc_{1}\Xc_{1} + \Qc_{2}\Xc_{2}.
\end{equation}
Suppose that input values are $\Xc_{1} = \alpha^2, \Xc_{2} = \alpha$ and channel coefficients are
$\Qc_{1}=\alpha, \Qc_{2}=1$. The received signal is obtained by
\begin{equation}
\Yc=\Qc_{1}\Xc_{1} + \Qc_{2}\Xc_{2} = \alpha^2.
\end{equation}
Equivalently, we have:
\begin{eqnarray*}
\Phi(\Yc) &=& \Gamma(\Qc_{1})\Phi(\Xc_{1}) + \Gamma(\Qc_{2})\Phi(\Xc_{2})\\
&=& \left[
  \begin{array}{cc}
    0 & 1 \\
    1 & 1 \\
  \end{array}
\right]\left[
         \begin{array}{c}
           1 \\
           1 \\
         \end{array}
       \right] + \left[
  \begin{array}{cc}
    1 & 0 \\
    0 & 1 \\
  \end{array}
\right]\left[
         \begin{array}{c}
           0 \\
           1 \\
         \end{array}
       \right]=\left[
         \begin{array}{c}
           1 \\
           1 \\
         \end{array}
       \right].
\end{eqnarray*}\hfill $\lozenge$
\end{example}

\section{System Models}\label{sec:SM}

In this paper we study interference management schemes such as interference alignment (IA)  and aligned network diagonalization (AND),
originally developed for Gaussian interference networks, when the channel is defined over a finite-field and the coefficients have particular algebraic structures.
For the case of Gaussian interference channels, the feasibility conditions of such schemes can be satisfied with arbitrarily high probability
provided that the channel coefficients are independently drawn from a continuous distribution \cite{Cadambe08,Gou,Shomorony1,Motahari}.
This is because the transmit signals and channel coefficients of Gaussian channels belong to the field of real numbers $\RR$ or complex numbers $\CC$, both of which are ``dense" and the set of coefficients for which matrices do not satisfy feasibility conditions have zero measure.
In the finite-field counterparts, however, signals and channel coefficients belong to a finite-field and hence the feasibility conditions
do not hold, in general, with probability 1. In this work we formulate the feasibility conditions of such schemes by taking the algebraic
structures of channel matrices into account and derive the probability that the feasibility conditions hold.
Throughout the paper, we consider the following three models:
\begin{description}
\item[1)] \textbf{Scalar $m$-th extension field models}: Each channel use is a symbol of $\FF_{p^m}$, i.e., transmit signals, channel coefficients, and received signals belong to $\FF_{p^m}$. The channel coefficients can be either time-invariant or time-varying. However, in this paper we focus on the time-invariant case, i.e., the channel coefficients are fixed over the whole transmission duration.
\item[2)] \textbf{Scalar ground-field models}: Each channel use is a symbol of the ground-field $\FF_{p}$, i.e., transmit signals, channel coefficients, and received signals belong to $\FF_{p}$.\footnote{For completeness, we should have considered a general ``ground'' field $\FF_q$ with $q$ equal to a prime power. 
Nevertheless, we found that restricting to $\FF_p$ makes exposition simpler, and allows a simple comparison of the various models
when $m$ channel uses of the scalar ground field model correspond to one channel use of the $m$-th extension field model. Of course, all conclusions
obtained for the scalar ground field models over $\FF_p$ carry on to a general $\FF_q$ with prime power $q$, under the assumption that symbols of the extension field
$\FF_q$ cannot be treated as vectors over the corresponding ground field, i.e., that only linear operations in the field $\FF_q$ are allowed.} 
In this case, we consider time-varying coefficients, i.e., the coefficients change (randomly) at each channel use. This is motivated, for example, by a network implementing random linear network coding where the linear combination coefficients at the intermediate nodes are chosen at random \cite{Ho} and changed at every channel use. We assume that all coefficients are known a-priori to all transmitters and receivers. Notice that they can be known {\em causally}
at transmitters and perfectly at receivers, without changing the achievability results \cite{Cadambe08}.
\item[3)] \textbf{MIMO ground-field models}: Each channel use  consists of an $m$-dimensional vector over $\FF_{p}$, i.e., transmit signals and received signals belong to $\FF_{p}^m$ and the channel coefficients are $m \times m$ full-rank matrices over $\FF_{p}$. In this work we focus on the case of
time-invariant channel matrices.
\end{description}
In order to have a fair comparison of achievable rates, we measure rates in bits per $m$ ground field symbols. Hence, the capacity of a deterministic link sending a vector of $m$ ground field symbols (or, equivalently, a symbol of the extension field) is given by $m\log{p}$ bits per $m$ ground field symbols. For scalar ground field models, $m$ ground field symbols correspond to a super-symbol of $m$ channel uses in time. We refer to this vectorized channel as the $m$-symbol extension model, in contrast to the $m$-th extension field model and the $m$-dimensional MIMO model. Then all three models can be represented as MIMO channels with $m \times m$ channel matrices over $\FF_{p}$ while the structure of such matrices is generally different, depending on the model. The following definition provides the algebraic structure of the sets of $m\times m$ matrices over $\FF_{p}$ corresponding to the three vector models introduced above, and will be extensively used in the sequel.

\begin{definition}\label{def:channel} We let $\Mc(p,m)$ denote the set of all $m \times m$ matrices over $\FF_{p}$. The set of non-singular matrices of $\Mc(p,m)$ forms a multiplicative group, known as {\em general linear group}, denoted by $\mbox{GL}(p,m)$.
Also, we let $\Dc^{\star}(p,m)$ denote the set of $m \times m$ {\em invertible} diagonal matrices over $\FF_{p}$. We have:
$\Cc^{\star}(p,m) \subset \mbox{GL}(p,m) \subset \Mc(p,m)$ and $\Dc^{\star}(p,m) \subset \mbox{GL}(p,m) \subset \Mc(p,m)$.
Furthermore, both multiplicative groups $\Cc^{\star}(p,m)$ and $\Dc^{\star}(p,m)$ are commutative.
\hfill $\lozenge$
\end{definition}

Next, we define the types of networks examined in this paper.

\subsection{Two-Unicast Two-Hop Networks}\label{model:2-user}

\begin{figure}[ht]
\centerline{\includegraphics[width=8
cm]{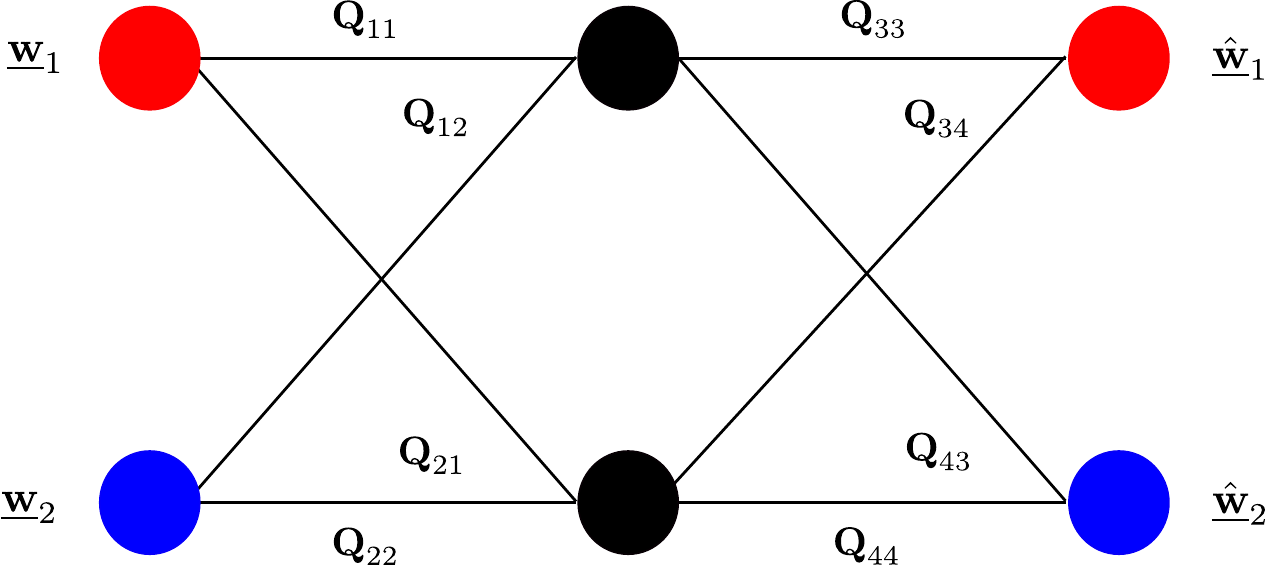}}
\caption{$2\times 2\times 2$ MIMO interference channel.}
\label{model222}
\end{figure}

\subsubsection{Scalar $m$-th extension field model}\label{model:field}
We consider a $2\times 2\times 2$ IC over $\FF_{p^m}$ where all nodes have a single input/output.
The first and second hops of the this channel are described by
\begin{equation}
\left[
  \begin{array}{c}
   \Yc_{1} \\
    \Yc_{2} \\
  \end{array}
\right] = \left[
            \begin{array}{cc}
              \Qc_{11} & \Qc_{12} \\
              \Qc_{21} & \Qc_{22} \\
            \end{array}
          \right]\left[
  \begin{array}{c}
    \Xc_{1} \\
   \Xc_{2} \\
  \end{array}
\right]\label{model:1-hop}
\end{equation}
and by
\begin{equation}
\left[
  \begin{array}{c}
    \Yc_{3} \\
    \Yc_{4} \\
  \end{array}
\right] = \left[
            \begin{array}{cc}
              \Qc_{33} & \Qc_{34} \\
              \Qc_{43} & \Qc_{44} \\
            \end{array}
          \right]\left[
  \begin{array}{c}
   \Xc_{3} \\
    \Xc_{4} \\
  \end{array}
\right], \label{model:2-hop}
\end{equation}
respectively, where $\Xc_{k} \in \FF_{p^m}, k=1,2,3,4$ and $\Yc_{\ell} \in \FF_{p^m}, \ell=1,2,3,4$.
Here, the channel coefficients $\Qc_{\ell k} \in \FF_{p^{m}}^{\star}$ are fixed and known to all nodes.
This model will be studied in Section~\ref{sec:FE}.

\subsubsection{Scalar ground-field model}\label{model:symbol}

We consider a $2\times 2\times 2$ IC over $\FF_{p}$ where all nodes have a single input/output.
As noticed before, each transmitter/receiver performs the coding/decoding over $m$ channel uses (i.e., $m$ symbol-extension).
The first and second hops of the this channel are described by
\begin{equation}
\left[
  \begin{array}{c}
   y_{1}[t] \\
   y_{2}[t] \\
  \end{array}
\right] = \left[
            \begin{array}{cc}
              q_{11}[t] & q_{12}[t] \\
              q_{21}[t] & q_{22}[t] \\
            \end{array}
          \right]\left[
  \begin{array}{c}
    x_{1}[t] \\
   x_{2}[t] \\
  \end{array}
\right]
\end{equation}
and by
\begin{equation}
\left[
  \begin{array}{c}
   y_{3}[t] \\
   y_{4}[t] \\
  \end{array}
\right] = \left[
            \begin{array}{cc}
              q_{33}[t] & q_{34}[t] \\
              q_{43}[t] & q_{44}[t] \\
            \end{array}
          \right]\left[
  \begin{array}{c}
    x_{3}[t] \\
   x_{4}[t] \\
  \end{array}
\right],
\end{equation}
respectively, where $x_{k}[t] \in \FF_{p}, k=1,2,3,4$, $y_{\ell}[t] \in \FF_{p}, \ell=1,2,3,4$, and the time-varying channel
coefficients $q_{\ell k}[t] \in \FF_{p}^{\star}$. This model will be studied in Section~\ref{sec:FE}.

\subsubsection{MIMO ground-field model}\label{model:MIMO}

We consider a $2\times 2\times 2$ IC over $\FF_{p}$ where all nodes have $m$ multiple inputs/outputs.
The first and second hops of the this channel are described by
\begin{equation}
\left[
  \begin{array}{c}
   \yv_{1} \\
    \yv_{2} \\
  \end{array}
\right] = \left[
            \begin{array}{cc}
              \Qm_{11} & \Qm_{12} \\
              \Qm_{21} & \Qm_{22} \\
            \end{array}
          \right]\left[
  \begin{array}{c}
    \xv_{1} \\
   \xv_{2} \\
  \end{array}
\right]
\end{equation}
and by
\begin{equation}
\left[
  \begin{array}{c}
    \yv_{3} \\
    \yv_{4} \\
  \end{array}
\right] = \left[
            \begin{array}{cc}
              \Qm_{33} & \Qm_{34} \\
              \Qm_{43} & \Qm_{44} \\
            \end{array}
          \right]\left[
  \begin{array}{c}
   \yv_{3} \\
    \yv_{4} \\
  \end{array}
\right],
\end{equation}
respectively, where $\xv_{k} \in \FF_{p}^{m}, k=1,2,3,4$ and $\yv_{\ell} \in \FF_{p}^{m}, \ell=1,2,3,4$.
Here, the channel coefficients $\Qm_{\ell k} \in \mbox{GL}(p,m)$ are fixed and known to all nodes.
This model will be studied in Section~\ref{sec:MIMO}.

\begin{remark}\label{remark:channel} (\textbf{MIMO Representation}) The scalar models in Sections~\ref{model:field} and~\ref{model:symbol} can be represented by MIMO models where all nodes have $m$ multiple inputs/outputs. In this case, the $m \times m$ channel matrices are defined as
\begin{itemize}
\item Scalar $m$-th extension field model:
\begin{equation}
\Qm_{\ell k} = \Gamma(\Qc_{\ell k}) \in \Cc^{\star}(p,m)
\end{equation}
\item Scalar ground-field model ($m$ channel uses):
\begin{equation}
\Qm_{\ell k} = \left[
                 \begin{array}{ccc}
                   q_{\ell k} [1] &  &  \\
                    & \ddots &  \\
                    &  & q_{\ell k} [m] \\
                 \end{array}
               \right] = \diag(q_{\ell k}[1],\ldots,q_{\ell k}[m]) \in \Dc^{\star}(p,m).
\end{equation}
\end{itemize}
All three models can be represented as in  Fig.~\ref{model222}.
However, they have different algebraic structures of their channel matrices,
such that $\Qm_{\ell k}\in \Cc^{\star}(p,m)$ for field-extension model,
$\Qm_{\ell k}\in \Dc^{\star}(p,m)$ for ground-field symbol-extension model, and
$\Qm_{\ell k}\in \mbox{GL}(p,m)$ for the ground-field MIMO model.
\hfill $\lozenge$
\end{remark}

\subsection{Three-User MIMO IC}\label{model:3-user}

Similarly to Section~\ref{model:2-user}, we consider 3-user MIMO IC with $m$ multiple inputs/outputs and with different structures of channel matrices.
The general channel model is given by
\begin{equation}
\yv_{\ell} = \Qm_{\ell 1}\xv_{1} + \Qm_{\ell 2}\xv_{2} + \Qm_{\ell 3}\xv_{3}
\end{equation}
where $\xv_{k} \in \FF_{p}^{m}, k=1,2,3$, $\yv_{\ell} \in \FF_{p}^{m}, \ell=1,2,3$, and $\Qm_{\ell k} \in \FF_{p}^{m \times m}$, $k,\ell=1,2,3$.
In this paper we consider three different algebraic structures for the channel matrices, namely: 1) $\Qm_{\ell k} \in \Cc^{\star}(p,m)$ for the
field-extension model;  2) $\Qm_{\ell k} \in \Dc^{\star}(p,m)$ for the ground-field model with symbol extension;
3) $\Qm_{\ell k} \in \mbox{GL}(p,m)$ for the MIMO ground-field model.

\section{Two-Unicast Two-Hop Scalar IC over $\FF_{p^m}$}\label{sec:FE}

The main result of this section is:

\begin{theorem}\label{thm:222-Field} For the $2\times 2\times 2$ IC over $\FF_{p^m}$,  the sum-rate of $(2m-1)\log{p}$ is achievable if $\deg(\pi_{\gamma}(x)) = m$ and $\deg(\pi_{\gamma'}(x)) = m$ ({\em feasibility conditions}),  where
\begin{eqnarray}
\gamma = \Qc_{11}^{-1}\Qc_{12}\Qc_{22}^{-1}\Qc_{21}\mbox{ and }\gamma' = \Qc_{33}^{-1}\Qc_{34}\Qc_{44}^{-1}\Qc_{43}\label{eq:cond}.
\end{eqnarray}
\end{theorem}
\begin{IEEEproof}
See Section~\ref{sec:AND}.
\end{IEEEproof}
Also, we obtain the normalized achievable sum-rate with respect to the interference-free channel capacity $m\log{p}$ of a single link,
when either $m$ or $p$ goes to infinity.  This metric is analogous to the sum degrees-of-freedom (sum-DoF) for Gaussian channels.

\begin{theorem}\label{thm2} If the channel coefficients $\Qc_{\ell k}$ are independently and uniformly drawn from $\FF_{p^m}^{\star}$, the following normalized sum-rates are achievable with probability $1$:
\begin{eqnarray}
d_{\rm{sum}}(p)&=&\lim_{m\rightarrow \infty}\frac{R_{\rm{sum}}(p,m)}{m\log{p}} =  2\label{eq:res1}\\
d_{\rm{sum}}(m)&=&\lim_{p \rightarrow \infty}\frac{R_{\rm{sum}}(p,m)}{m\log{p}} = \frac{2m-1}{m}\label{eq:res2}
\end{eqnarray}where $R_{\rm{sum}}(p,m)$ denotes the achievable sum-rate for given finite-field $\FF_{p^m}$.
\end{theorem}
\begin{IEEEproof} The proof consists of showing that the feasibility conditions in Theorem~\ref{thm:222-Field} are satisfied with probability $1$ in the limits. Let $N(p,m)$ denote the number of {\em monic irreducible polynomials} of degree-$m$ over $\FF_{p}$. From \cite[Th. 15]{Mac}, we have:
\begin{equation*}
N(p,m) = \frac{1}{m} \sum_{d|m} \nu(d)p^{m/d} \label{eq:irr}
\end{equation*} where $\nu(d)$ denotes the M\"{o}bius function, defined by
\begin{equation*}
\nu(d) = \left\{
           \begin{array}{ll}
             1 & \hbox{if } d=1 \\
             (-1)^{r}, & \hbox{if $d$ is the product of $r$ distinct primes} \\
             0, & \hbox{otherwise.}
           \end{array}
         \right.
\end{equation*} From the basic finite-field theory, we have two results: 1) Each degree-$m$ monic irreducible polynomial has $m$ distinct roots in $\FF_{p^m}$ and is a degree-$m$ minimal polynomial of such roots (see \cite[Cor. 1.3.4]{Niederreiter}). For example, if $\alpha$ is a root of $f(x)$ then $\alpha,\alpha^{p},\ldots,\alpha^{p^{m-1}}$ are $m$ distinct roots of $f(x)$. 2) If $f(x)$ and $g(x)$ are distinct monic irreducible polynomials over $\FF_{p}$ then $f(x)$ and $g(x)$ have no common roots in any extension of $\FF_{p}$ (see \cite[Cor. 3.1.5]{Robert}). Thus, we have $mN(p,m)$ distinct elements in $\FF_{p^m}$ with degree-$m$ minimal polynomial. Also, we can derive a simple lower-bound on  $mN(p,m)$ by setting $\nu(d) = -1$ for any  $d$ with $d|m, d > 1$:
\begin{equation*}
m N(p,m) \geq  p^{m} - \sum_{d|m, d>1}p^{m/d}.
\end{equation*} Using this bound and the fact that $\gamma$ defined in (\ref{eq:cond}), is uniformly distributed over $\FF_{p^m}^{\star}$, we have:
\begin{eqnarray}
\PP(\{\deg(\pi_{\gamma}(x))=m\}) &=& \frac{mN(p,m)}{p^{m}-1} \label{eq:prob1}\\
&\geq& \frac{p^{m} - \sum_{d|m, d>1}p^{m/d}}{p^{m}}\nonumber\\
&=& 1- \sum_{d|m, d>1}p^{m(1/d -1)}\nonumber \\
 &\stackrel{(a)}{\geq}& 1-\sum_{d|m, d>1} p^{(-m/2)} \nonumber \\
&\geq& 1- p^{(-m/2)}\sum_{d=1}^{m} \nonumber \\
&=& 1 - \frac{m(m+1)}{2p^{m/2}} \nonumber \\
&\rightarrow& 1 \mbox{ if either $m$ or $p$ goes to infinity} \nonumber
\end{eqnarray} where (a) is due to the fact that $ -1/2 \geq 1/d -1$ for $d\geq 2$. With the same procedures, we can also prove that $\PP(\{\deg(\pi_{\gamma'}(x))=m\})$ goes to 1 if either $m$ or $p$ approaches to infinity. This completes the proof.
\end{IEEEproof}

From Theorem~\ref{thm2}, we notice that the optimal sum-capacity is achieved for any prime $p$ as long as $m$ goes
to infinity. Using (\ref{eq:prob1}) and the fact that $\gamma$ and $\gamma'$ are independently and uniformly distributed
over $\FF^\star_{p^m}$, the probability that the feasibility conditions of Theorem~\ref{thm:222-Field} holds is given by:
\begin{equation}\label{probf:C}
\PP_{\rm{FE}} = \left(\frac{p^{m} + \sum_{d|m, d>1}\nu(d)p^{m/d}}{p^{m}-1}\right)^2.
\end{equation} In particular when $m$ is a prime, this probability takes on the simple form
\begin{equation}
\PP_{\rm{FE}} = \left(\frac{p^{m} - p}{p^{m} - 1}\right)^2.\label{FE:prime}
\end{equation}
\begin{figure*}
\centerline{\includegraphics[width=12cm]{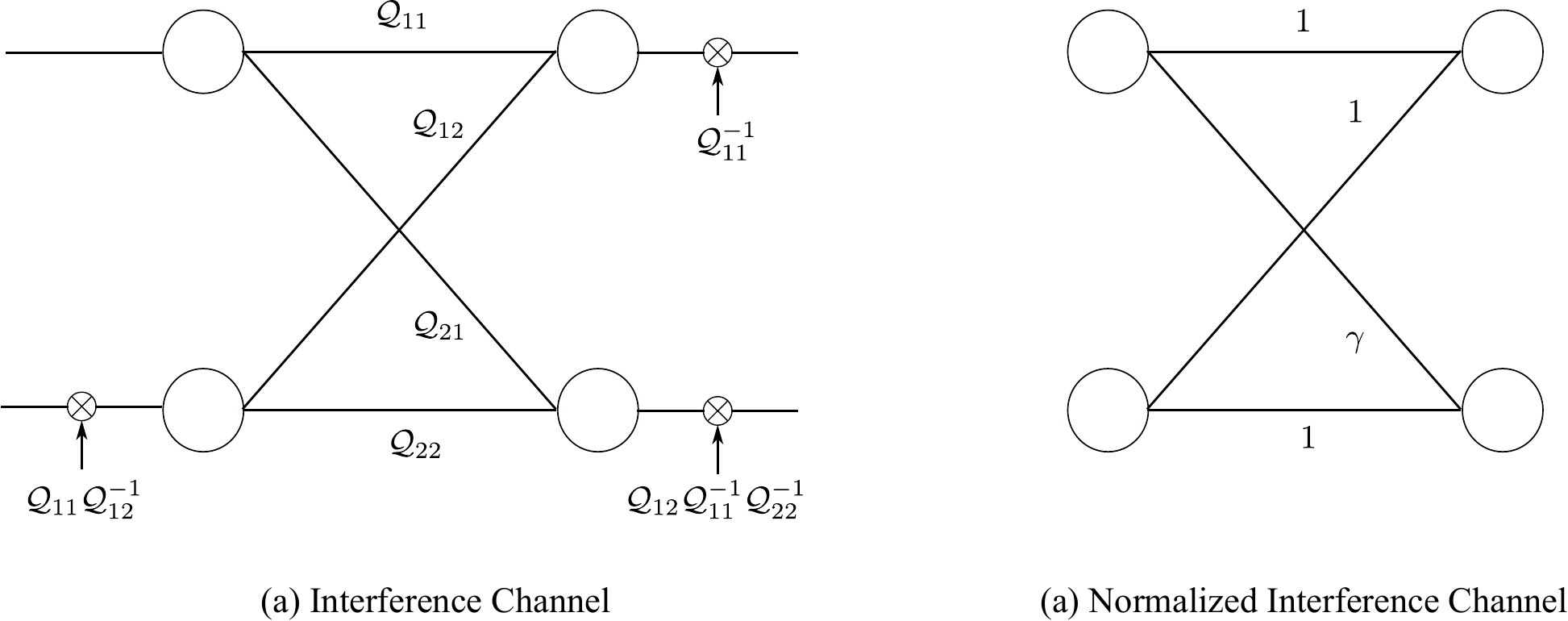}}
\caption{Normalized Interference Channel with non-zero coefficients where $\gamma$ is defined in Theorem~\ref{thm:222-Field}.}
\label{normalizedIC}
\end{figure*}

\begin{remark} It is interesting to investigate the network structures for which the feasibility condition in Theorem~\ref{thm:222-Field} does not hold.
Without loss of generality, we consider normalized channel coefficients by invertible operations at transmitters and receivers,
as done in \cite{Krishnamurthy} (see Fig.~\ref{normalizedIC}).
In the resulting normalized model, the channel coefficients are $\Qc_{11}=\Qc_{12}=\Qc_{22}=1$ and $\Qc_{21} = \gamma$.
For ease of exposition, we consider that $m$ is a prime.
Then, if $\pi_{\gamma}(x), \pi_{\gamma'}(x) < m$, we have that $\gamma, \gamma' \in \FF_{p}$ since the degree of minimal polynomial
should divide $m$. In this case, the transformed $2\times 2\times 2$ MIMO IC over $\FF_{p}$ can be decomposed into $m$ parallel
$2\times 2\times 2$ scalar ICs over the ground-field $\FF_{p}$ since all channel matrices are either identity matrix or scaled identity matrix.
The capacity of this channel model is still unknown.\hfill $\lozenge$
\end{remark}

\begin{figure}[ht]
\centerline{\includegraphics[width=14cm]{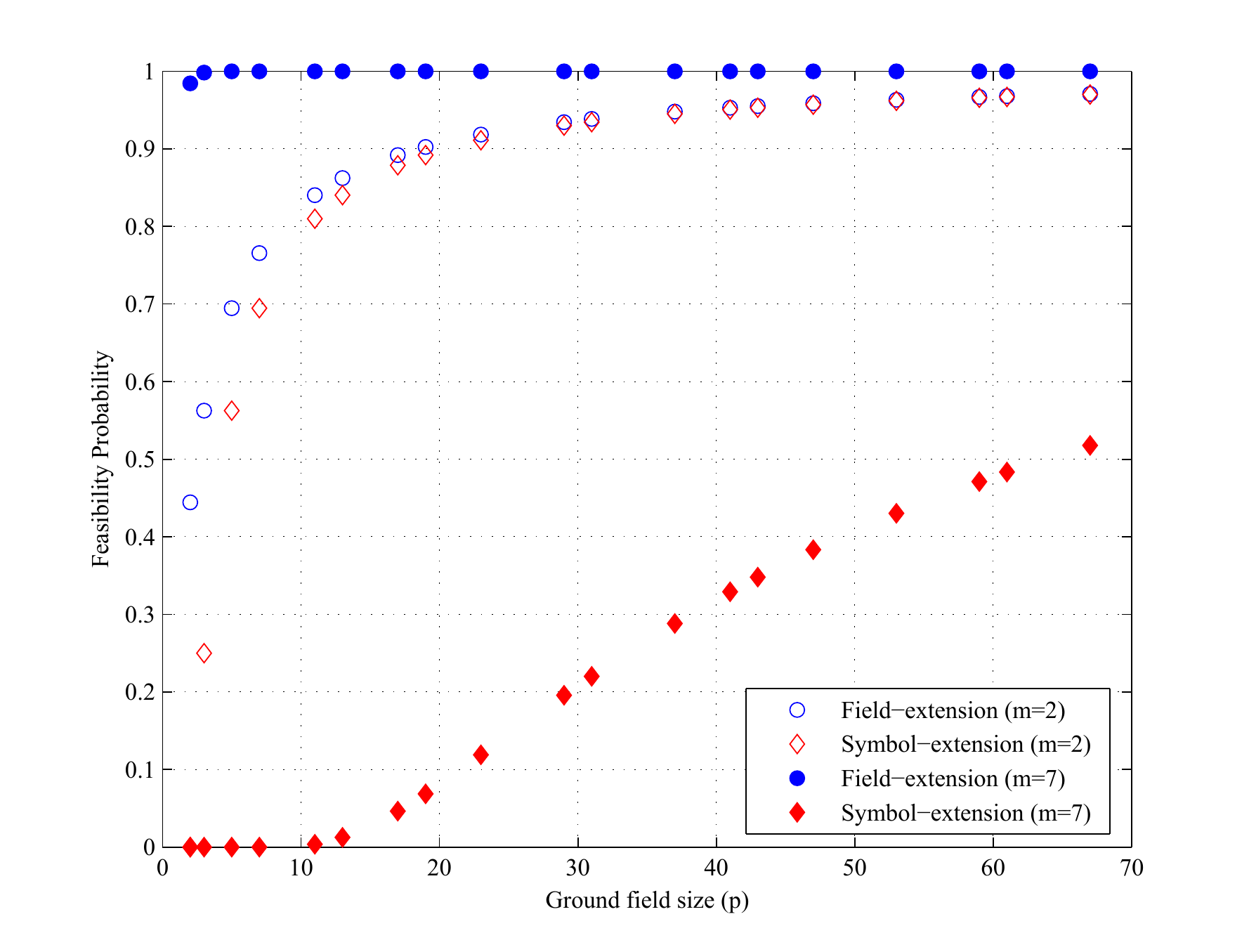}}
\caption{The comparison between symbol-extension and field-extension with respect to feasibility probability.}
\label{comp}
\end{figure}

\subsection{Comparison with $m$-symbol extension}\label{sec:symbol}

In this section we compare the previous result with the case of an $m$-symbol extension $2\times 2\times 2$ IC over $\FF_{p}$, with time-varying channel coefficients (see model in Section~\ref{model:symbol}). In this case, the MIMO $2\times 2\times 2$ IC has $m\times m$ channel matrices in the form:
\begin{equation}
\Qm_{\ell k} = \diag(q_{\ell k}[1],\ldots,q_{\ell k}[m]).
\end{equation} One may expect that two MIMO channel models (namely, the one obtained by field-extension and the other by symbol-extension) are equivalent,
since both of them have $\approx p^m$ possible channel matrices, and these matrices form commutative groups.
For the scalar ground-field model with $m$-fold symbol extension, the same achievable scheme of Section~\ref{sec:AND} can be applied with
different feasibility conditions, namely, that the diagonal elements of the products of channel matrices,
defined by $\Qm=\Qm_{11}^{-1}\Qm_{12}\Qm_{22}^{-1}\Qm_{21}$, are distinct \cite{Gou}. The probability of distinct non-zero elements for one matrix
is given by
\begin{equation}
1 \times \frac{p-2}{p-1} \times \frac{p-3}{p-1} \times \cdots\times \frac{p-m}{p-1} = (p-1)^{-m}\prod_{i=1}^{m}(p-i).
\end{equation}
Since the matrices in the first and second hop are independent,  the feasibility probability of the scalar ground-field model with $m$-fold symbol extension is given by
\begin{equation}\label{probf:D}
\PP_{\rm{SE}} = \left((p-1)^{-m}\prod_{i=1}^{m}(p-i)\right)^2,
\end{equation}
Notice that this probability is equal to zero when $m \geq p$. Hence, a necessary feasibility condition is  $p > m$.
The probability $\PP_{\rm{SE}}$  goes to 1, as for the case of field-extension, when $p \rightarrow \infty$.
However, the probability goes to zero when $m \rightarrow \infty$ and $p$ is fixed,
while we have seen before that in the field-extension the feasibility probability goes to 1 also in this case.
From a numerical result viewpoint, it is interesting to observe that the feasibility probability of the extension-field model 
is very close to $1$ for any $p$ when $m$ is a prime larger or equal to 7 (see Fig.~\ref{comp} and eq. (\ref{FE:prime})).

\subsection{Proof of Theorem 1: Achievable scheme}\label{sec:AND}

We prove Theorem~\ref{thm:222-Field} using AND, under the assumption that $\gamma$ and $\gamma'$ have degree-$m$ minimal polynomial. From Section~\ref{sec:MAC}, we can transform the $2\times 2\times 2$ scalar IC over $\FF_{p^m}$ in (\ref{model:1-hop}) and (\ref{model:2-hop}) into MIMO IC over $\FF_{p}$ with channel coefficients  $\Qm_{\ell k}= \Gamma(\Qc_{\ell k}) \in \Cc^{\star}(p,m)$. Notice that $\Qm_{\ell k}$ is always full rank over $\FF_{p}$. The proposed coding scheme is performed for the transformed MIMO channel and the one-to-one mapping $\Phi(\cdot)$ is used to transmit  the coded messages via the original channels.  In order to transmit $(2m-1)$ streams,  source $1$ sends $m$ independent messages
$\{w_{1,\ell} \in \FF_{p}: \ell=1,\ldots,m\}$ to destination 1 and source 2 sends $m-1$ independent messages $\{w_{2,\ell} \in \FF_{p}: \ell=1,\ldots,m-1\}$ to destination 2. For simplicity, we also use the vector representation of messages as $\wv_{1} = [w_{1,1},\ldots,w_{1,m}]^{\transp}$ and $\wv_{2} = [w_{2,1},\ldots,w_{2,m-1}]^{\transp}$.

\subsubsection{Encoding at the sources}

We let $\Vm_{1} = [\vv_{1,1},\ldots,\vv_{1,m}] \in \FF_{p}^{m \times m}$ and $\Vm_{2} = [\vv_{2,1},\ldots,\vv_{2,m-1}] \in \FF_{p}^{m \times m-1}$ denote the precoding matrices used at sources 1 and 2, respectively, chosen to satisfy the {\em alignment conditions}:
\begin{eqnarray}
\Qm_{11}\vv_{1,\ell+1} &=& \Qm_{12}\vv_{2,\ell}\nonumber\\
\Qm_{21}\vv_{1,\ell} &=& \Qm_{22}\vv_{2,\ell}\label{eq:alignment}
\end{eqnarray}for $\ell=1,\ldots,m-1$. For alignment, we use the construction method proposed in \cite{Gou}:
\begin{eqnarray}
\vv_{1,\ell+1} &=& (\Qm_{11}^{-1}\Qm_{12}\Qm_{22}^{-1}\Qm_{21})^{\ell}\vv_{1,1} \label{const1}\\
\vv_{2,\ell} &=& (\Qm_{22}^{-1}\Qm_{21}\Qm_{11}^{-1}\Qm_{12})^{\ell-1}\Qm_{22}^{-1}\Qm_{21}\vv_{1,1}\label{const2}
\end{eqnarray}for $\ell=1,\ldots,m-1$. Using $\Gamma(\cdot)$ and $\gamma$ defined in (\ref{eq:cond}), the above constructions can be rewritten as
\begin{eqnarray}
\vv_{1,\ell+1} &=&  \Gamma((\Qc_{11}^{-1}\Qc_{12}\Qc_{22}^{-1}\Qc_{21})^{\ell})\vv_{1,1} = \Gamma(\gamma^{\ell})\vv_{1,1}\label{eq:const1}\\
\vv_{2,\ell} &=&\Gamma(\Qc_{22}^{-1}\Qc_{21})\Gamma(\gamma^{\ell-1})\vv_{1,1}\label{eq:const2}
\end{eqnarray}for $\ell=1,\ldots,m-1$.

Source $k$ precodes its message as $\xv_{k}=\Vm_{k}\wv_{k}$
and produces the channel input
\begin{equation}
\Xc_{k} = \Phi^{-1}(\xv_{k}) \in \FF_{p^m}, \;\;\; k=1,2.
\end{equation} Then, $\Xc_{1}$ and $\Xc_{2}$ are transmitted over channels.

\begin{remark}
Notice that the precoding has to be performed over the ground field $\FF_p$, since the multiplication $\Vm_k \wv_k$ does not correspond to
the multiplication of $\Wc_k$ by an element of $\FF_{p^m}$ corresponding to $\Vm_k$.
This is because, in general,  $\Vm_k$ is not a power of the companion matrix $\Cm$.  \hfill $\lozenge$
\end{remark}

\subsubsection{Relaying operations}

Relays decode linear combinations of source messages and forward the precoded linear combinations to destination as follows.

Relay 1 observes:
\begin{eqnarray*}
\Yc_{1} = \Qc_{11}\Xc_{1} + \Qc_{12}\Xc_{2} \in \FF_{p^m}
\end{eqnarray*} and maps the received signal onto ground field $\FF_{p}$:
\begin{eqnarray}
\Phi(\Yc_{1}) &=& \Qm_{11}\Phi(\Xc_{1})+ \Qm_{12}\Phi(\Xc_{2})\nonumber\\
&=& \Qm_{11}\Vm_{1}\wv_{1} + \Qm_{12}\Vm_{2}\wv_{2}\nonumber\\
&\stackrel{(a)}{=}&\Qm_{11}\Vm_{1} \underbrace{\left[
                     \begin{array}{c}
                       w_{1,1} \\
                       w_{1,2} + w_{2,1} \\
                       \vdots \\
                       w_{1,m} + w_{2,m-1} \\
                     \end{array}
                   \right]}_{\eqdef \uv_{1}} \label{received1}
\end{eqnarray} where (a) is due to the fact that precoding vectors satisfy the alignment conditions in (\ref{eq:alignment}). Since $\Vm_{1}$ is full-rank over $\FF_{p}$ by Lemma~\ref{lem:full}, relay 1 can decode $\uv_{1}$ (i.e., linear combinations of source messages).

Similarly, relay 2 observes the aligned signals over $\FF_{p}$:
\begin{eqnarray}
\Phi(\Yc_{2}) &=& \Qm_{21}\Phi(\Xc_{1}) + \Qm_{22}\Phi(\Xc_{2})\nonumber\\
&=& \Qm_{21}\Vm_{1}\wv_{1} + \Qm_{22}\Vm_{2}\wv_{2}\nonumber\\
&\stackrel{(a)}{=}&\Qm_{21}\Vm_{1}\underbrace{ \left[
                     \begin{array}{c}
                       w_{1,1} + w_{2,1} \\
                       \vdots \\
                       w_{1,m-1} + w_{2,m-1} \\
                       w_{1,m}\\
                     \end{array}
                   \right]}_{\eqdef \uv_{2}}\label{received2}
\end{eqnarray}where (a) is due to the fact that precoding vectors satisfy the alignment conditions in (\ref{eq:alignment}). Since $\Vm_{1}$ is full-rank over $\FF_{p}$ by Lemma~\ref{lem:full}, relay 2 can decode  $\uv_{2}$.


\begin{lemma}\label{lem:full} Assume that $\deg(\pi_{\gamma}(x)) = m$. Then, by choosing $\vv_{1,1} = \Phi(1)$, the matrix $\Vm_{1}$ has full rank.
\end{lemma}
\begin{IEEEproof} Using $\vv_{1,1} = \Phi(1)$,  we have:
\begin{equation}
\Gamma(\gamma^{\ell}) \vv_{1,1} = \Phi(\Gamma^{-1}(\Gamma(\gamma^{\ell}))\Phi^{-1}(\vv_{1,1})) = \Phi(\gamma^{\ell})\label{eq:form}.
\end{equation} From (\ref{eq:const1}) and  (\ref{eq:form}), the precoding matrix $\Vm_{1}$ can be written as
\begin{eqnarray*}
\Vm_{1} &=& [\vv_{1,1},\ldots,\vv_{1,m}] \\
&=&[\Phi(1), \Phi(\gamma), \Phi(\gamma^2),\ldots, \Phi(\gamma^{m-1})].\label{eq:V1}
\end{eqnarray*}  Since $\gamma$ is assumed to have degree-$m$ minimal polynomial, the following holds:
\begin{equation*}
b_{0} + b_{1}\gamma + \cdots + b_{m-1}\gamma^{m-1} \neq \zerov
\end{equation*} for any non-zero coefficients vector $(b_{0},\ldots,b_{m-1}) \in \FF_{p}^{m}$. Using this, we can prove that $\Vm_{1}$ has $m$ linearly independent columns:
\begin{eqnarray*}
&&b_{0}\Phi(1)+ b_{1}\Phi(\gamma) + \cdots + b_{m-1}\Phi(\gamma^{m-1}) \\
&=&\Phi(b_{0}) + \Phi(b_{1}\gamma) +  \cdots + \Phi(b_{m-1}\gamma^{m-1})\\
&=& \Phi(b_{0} + b_{1}\gamma + \cdots + b_{m-1}\gamma^{m-1})\neq \textbf{0}
\end{eqnarray*}  for any non-zero coefficients vector $(b_{0},\ldots,b_{m-1})  \in \FF_{p}^{m}$. This completes the proof.
\end{IEEEproof}

At this point, relay 1 precodes the decoded linear combinations as $\xv_{3} = \Sm_{11}\Vm_{3}\uv_{1}$  and produces the channel input
\begin{equation}
\Xc_{3}=\Phi^{-1}(\xv_{3}) \in \FF_{p^{m}}
\end{equation} Likewise, relay 2 precodes the decoded linear combinations as $\xv_{4} = \Sm_{21}\Vm_{3}\uv_{2}$ and produces the channel input
\begin{equation}
\Xc_{4}=\Phi^{-1}(\xv_{4}) \in \FF_{p^{m}}
\end{equation} where
\begin{equation}
\Sm=\left[
       \begin{array}{cc}
         \Sm_{11} & \Sm_{12} \\
         \Sm_{21} & \Sm_{22} \\
       \end{array}
     \right] =\left[
       \begin{array}{cc}
         \Qm_{33} & \Qm_{34} \\
         \Qm_{43} & \Qm_{44} \\
       \end{array}
     \right]^{-1}\label{eq:S}
\end{equation}  and $\Vm_{3}$ are chosen to satisfy the alignment conditions in (\ref{eq:alignment}) with respect to $\Sm$:
\begin{eqnarray}
\vv_{3,\ell+1} &=& \Gamma(\gamma'^{\ell})\vv_{3,1}\label{eq:const3}\\
\vv_{4,\ell} &=&\Gamma(\Sc_{22}^{-1}\Sc_{21})\Gamma((\Sc_{11}^{-1}\Sc_{12}\Sc_{22}^{-1}\Sc_{21})^{\ell-1})\vv_{3,1}\nonumber\\
&=&\Gamma(\Sc_{22}^{-1}\Sc_{21})\Gamma(\gamma'^{\ell-1})\vv_{3,1}\label{eq:const4}
\end{eqnarray}for $\ell=1,\ldots,m-1$ where $\Sc_{ij} = \Gamma^{-1}(\Sm_{ij})$ and where $\gamma'$ is defined in (\ref{eq:cond}). Here, we used the fact that $\Sc_{11}^{-1}\Sc_{12}\Sc_{22}^{-1}\Sc_{21} = \Qc_{33}^{-1}\Qc_{34}\Qc_{44}^{-1}\Qc_{43}$. From Lemma~\ref{lem:full}, we can immediately prove that $\Vm_{3}$ and $\Vm_{4}$ are full rank by choosing $\vv_{3,1}=\Phi(1)$ since $\deg(\pi_{\gamma'}(x)) = m$. The other precoding vectors are completely determined by the (\ref{eq:const3}) and (\ref{eq:const4}).

From  (\ref{received1}) and (\ref{received2}), we can observe that the coefficients of the linear combinations only depend on alignment conditions, independent of channel coefficients. From this, we can produce the received signal for which the channel matrix is equal to the inverse of second-hop channel matrix. This is the key property to enable the network diagonalization. That is,
$\xv_{3}$ and $\xv_{4}$ are equal to received signals with channel coefficients $\Sm$:
\begin{eqnarray}
\left[
  \begin{array}{c}
   \xv_{3} \\
   \xv_{4} \\
  \end{array}
\right] &=&\left[
  \begin{array}{c}
    \Sm_{11}\Vm_{3}\uv_{1} \\
    \Sm_{21}\Vm_{3}\uv_{2} \\
  \end{array}
\right] = \left[
  \begin{array}{c}
    \Sm_{11}\Vm_{3} \wv_{1} + \Sm_{12}\Vm_{4}\wv_{2} \\
    \Sm_{21}\Vm_{3}\wv_{1} + \Sm_{22}\Vm_{4}\wv_{2} \\
  \end{array}
\right] \nonumber\\
&=& \left[
       \begin{array}{cc}
         \Qm_{33} & \Qm_{34} \\
         \Qm_{43} & \Qm_{44} \\
       \end{array}
     \right]^{-1}\left[
  \begin{array}{c}
    \Vm_{3}\wv_{1} \\
    \Vm_{3}\wv_{2} \\
  \end{array}
\right].\label{equiv}
\end{eqnarray}

\subsubsection{Decoding at the destinations}

Destinations 1 and 2 observe:
\begin{eqnarray*}
\left[
  \begin{array}{c}
    \Yc_{3} \\
    \Yc_{4} \\
  \end{array}
\right] = \left[
            \begin{array}{cc}
              \Qc_{33} & \Qc_{34} \\
              \Qc_{43} & \Qc_{44} \\
            \end{array}
          \right]\left[
  \begin{array}{c}
   \Xc_{3} \\
    \Xc_{4} \\
  \end{array}
\right].
\end{eqnarray*} By mapping the received signals onto the ground field $\FF_{p}$, we get:
\begin{eqnarray*}
\left[
  \begin{array}{c}
    \Phi(\Yc_{3}) \\
    \Phi(\Yc_{4}) \\
  \end{array}
\right]
&=&  \left[
            \begin{array}{cc}
              \Qm_{33} & \Qm_{34} \\
              \Qm_{43} & \Qm_{44} \\
            \end{array}
          \right] \left[
                              \begin{array}{c}
                              \Sm_{11}\Vm_{3}\uv_{1} \\
                                \Sm_{21}\Vm_{3}\uv_{2} \\
                              \end{array}
                            \right]\\
                            &\stackrel{(a)}{=}& \left[
                              \begin{array}{c}
                              \Vm_{3}\wv_{1} \\
                                \Vm_{4}\wv_{2} \\
                              \end{array}
                            \right]
\end{eqnarray*}  where (a) is due to the precoding at relays to satisfy (\ref{equiv}). This shows that destination 1 can decode $\wv_{1}$ using $\Vm_{3}^{-1}\Phi(\Yc_{3})$ and destination 2 can decode $\wv_{2}$ using $\Vm_{4}^{-1}\Phi(\Yc_{4})$. Hence, the proof is complete.

%
\section{Two-Unicast Two-Hop MIMO IC over $\FF_{p}$}\label{sec:MIMO}

We consider the channel model defined in Section~\ref{model:MIMO}. Recall that $\Qm_{\ell k} \in \FF_{p}^{m \times m}$ denotes the channel matrices between the source $k$ and the destination $\ell$ and they are assumed to be full rank over $\FF_{p}$, i.e., $\Qm_{\ell k} \in \mbox{GL}(p,m)$. Notice that they are neither diagonal matrices nor in the form of powers of companion matrix, and do not commute in general. Thus, it is not possible to straightforwardly apply the same approach developed before. Instead, we have to resort to {\em symbol-extension} by going to an extension field in order to obtain an AND scheme.

From (\ref{const1}) and (\ref{const2}),  a choice of the precoding matrix $\Vm_{1}$ that satisfies
the alignment conditions is given by:
\begin{equation}
\Vm_{1} = [\vv_{1,1},\Qm\vv_{1,1}\ldots,\Qm^{m-1}\vv_{1,1}]
\end{equation} where
\begin{equation}
\Qm = \Qm_{11}^{-1}\Qm_{12}\Qm_{22}^{-1}\Qm_{21}. \label{eq:defQ}
\end{equation}
We cannot make use of the result in Section~\ref{sec:AND} since $\Qm$ does not correspond to an element of $\FF_{p^m}$ through the matrix representation $\Gamma$. If $\Qm$ has $m$ distinct eigenvalues in $\FF_{p}$, then it has $m$ linearly independent eigenvectors.
Following \cite{Gou,Hong-J}, we can prove that $\Vm_{1}$ has full-rank if we choose $\vv_{1,1} = \Em\onev$ where $\Em$ consists of $m$ linearly independent eigenvectors of $\Qm$. In the case of the complex-valued Gaussian channel, we can always find $m$ distinct eigenvalues since the complex
field is algebraically closed (i.e., any polynomial over $\CC$ splits in $\CC$).
However, for the finite-field $\FF_{p}$, some eigenvalues of $\Qm$ may not exist in the ground-field $\FF_{p}$, depending on the
characteristic polynomial of $\Qm$ (denoted by $c(x)$). Suppose that this polynomial splits into irreducible factors  over $\FF_{p}$ in the form:
\begin{equation}
c(x)=\prod_{i=1}^{d}c_{i}(x)
\end{equation} where $\deg(c_{i}(x)) \geq \deg(c_{j}(x))$ if $i \leq j$ and $c_{i}(x) \in \FF_{p}[x]$ for all $i$. Let $\deg(c_{i}(x)) = r_{i}$.
If $r_{i} > 1$ then some eigenvalues of $\Qm$ do not exist in $\FF_{p}$. Also, since $c_{i}(x)$ is irreducible over $\FF_{p}$,  the ring of polynomial residues $L_{i}\eqdef\FF_{p}[x]/c_{i}(x)$ is isomorphic to $\FF_{p^{r_{i}}}$, the extension field of degree $r_{i}$ over $\FF_{p}$. From \cite[Cor. 1.3.4]{Niederreiter}, we have that $\FF_{p^{r_{i}}}$ is the {\em splitting field} of $c_{i}(x)$, i.e., it is the minimal field containing all roots of $c_{i}(x)$. From \cite[Th. 8]{Mac}, we know that $\FF_{p^s}$ contains a subfield $\FF_{p^k}$ if and only if $k$ divides $s$. Hence, $\FF_{p^r}$ contains all roots of $c(x)$ if $r$ is the {\em least common multiple} of $r_{i}$'s (i.e., $r=\mbox{lcm}(r_{1},\ldots,r_{d}))$. In short, $\FF_{p^r}$ is the splitting field of $c(x)$. Notice that an upper bound on the degree $r$ of such splitting field
is given by the maximum of the least common multiples over all possible integer partitions of the integer $m$.
Here, such integer partitions include the set of degrees of the irreducible factors of the characteristic polynomial $c(x)$, as well as many other
possibilities that do not correspond to actual factorizations (this is why we get an upper bound).
Unfortunately, to the best of the authors' knowledge, finding the maximum least common multiple over all integer partitions of a given integer $m$ seems to be an unsolved problem in number theory, and a general upper-bound on the degree of the splitting field is given trivially by $m!$ in \cite[Prop. 3.2.2]{Robert}.
However, this bound is quite loose as shown in the example of Table~\ref{table:upper}, where we computed
the upper-bound by finding exhaustively the least common multiple over all integer partitions of $m$.

\begin{table}
\caption{Upper-bound on required symbol-extension}
\begin{equation*}
\begin{array}{|c||ccccccccc|}
\hline
 m & 2 & 3 & 4 & 5 & 6 & 7 & 8 & 9 & 10 \\
  \hline
  \mbox{upper-bound}  & 2 & 3 & 4 & 6 & 6 & 12 & 15 & 20 & 30 \\
  \hline
  m! & 2 & 6 & 24 & 120 & 720 & 5040 & 40320 & 362880 & 3628800\\
  \hline
\end{array}
\end{equation*}
\label{table:upper}
\end{table}

\begin{figure}[ht]
\centerline{\includegraphics[width=16cm]{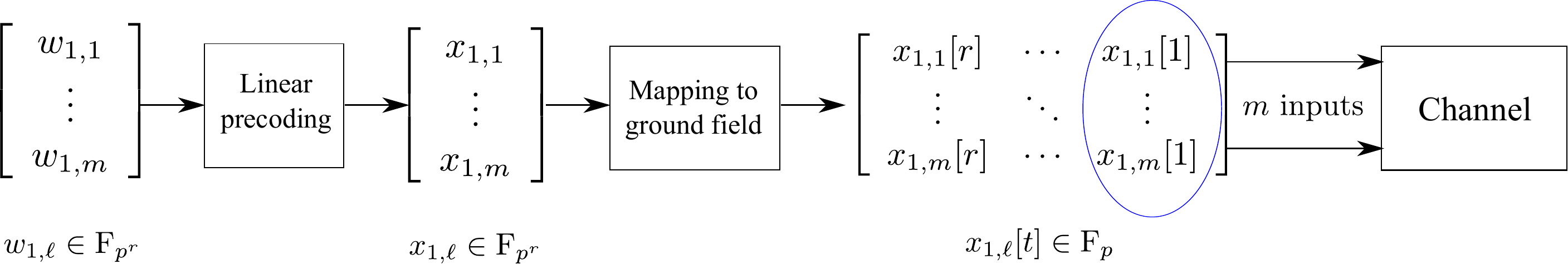}}
\caption{Encoding procedure over $r$ channel uses.}
\label{MIMO_Tx}
\end{figure}

Assume that $\Qm$ has $m$ distinct eigenvalues $\{\lambda_{i} \in \FF_{p^r}: i=1,\ldots,m\}$. From \cite[Th. 1.3.9]{Horn}, $\Qm$ is diagonalizable such as
\begin{equation}
\Qm = \Em\left[
           \begin{array}{ccc}
             \lambda_{1} &  &  \\
              & \ddots &  \\
              &  & \lambda_{m} \\
           \end{array}
         \right]\Em^{-1}
\end{equation}where the $i$-th column of $\Em$ is an eigenvector of $\Qm$ associated with $\lambda_{i}$. Choosing $\vv_{1,1} = \Em\onev \in \FF_{p^r}^{m}$, the alignment precoding matrix $\Vm_{1}$ can be rewritten as
\begin{eqnarray}
\Vm_{1} = \Em\underbrace{\left[
         \begin{array}{cccc}
           1 & \lambda_{1} & \cdots & \lambda_{1}^{m-1} \\
           \vdots & \vdots & \ddots & \vdots \\
           1 & \lambda_{m} & \cdots & \lambda_{m}^{m-1} \\
         \end{array}
       \right]}_{\eqdef \Jm}
\end{eqnarray} where $\Jm$ denotes the Vandermonde matrix. Therefore, the determinant of $\Vm_{1}$ is given by
\begin{eqnarray*}
\det(\Vm_{1}) &=& \det(\Em)\det(\Jm)\\
&=&\det(\Em)\prod_{1\leq i < j\leq m}(\lambda_{j} - \lambda_{i}) \neq 0.
\end{eqnarray*} This shows that we can find the precoding matrix $\Vm_{1}$ in the extension field $\FF_{p^r}$ that satisfies the alignment condition and is full rank over $\FF_{p^r}$. Accordingly, encoding and decoding should be performed over the extension field as shown in Fig.~\ref{MIMO_Tx}.

For completeness, we present the resulting
coding scheme over the $r$-symbol extension (i.e., over $r$ time slots).
Source 1 precodes its message $\wv_{1} \in \FF_{p^r}^{m}$ using precoding matrix $\Vm_{1} \in \FF_{p^r}^{m\times m}$:
\begin{equation*}
\xv_{1} = \Vm_{1}\wv_{1} \in \FF_{p^r}^{m}
\end{equation*}  and transmits the $t$-th column of $\Xm_{1}=\Phi^{\transp}(\xv_{1}) \in \FF_{p}^{m \times r}$ at time slot $t$ for $t=1,\ldots,r$ where $\Phi^{\transp}: \FF_{p^r} \rightarrow [\FF_{p},\ldots,\FF_{p}]$ (notice that differently from (\ref{def:Phi}), it maps the elements of $\FF_{p^r}$ to the $r$-dimensional row vectors). Similarly, source 2 precodes its message $\wv_{2} \in \FF_{p^r}^{m-1}$ using precoding matrix $\Vm_{2} \in \FF_{p^r}^{m\times m-1}$:
\begin{equation*}
\xv_{2} = \Vm_{2}\wv_{2} \in \FF_{p^r}^{m}
\end{equation*}  and transmits the $t$-th column of $\Xm_{2}=\Phi^{\transp}(\xv_{2}) \in \FF_{p}^{m \times r}$ at time slot $t$ for $t=1,\ldots,r$.

After $r$ time slots, relays perform the decoding procedures as follows. Relay 1 observes:
\begin{eqnarray*}
\Ym_{1} &=& \Qm_{11}\Xm_{1} + \Qm_{12}\Xm_{2} \in \FF_{p}^{m \times r}.
\end{eqnarray*} Let $\Phi^{-\transp}(\cdot)$ denote the inverse function of $\Phi^{\transp}(\cdot)$, i.e., $\Phi^{-\transp}: [\FF_{p},\ldots,\FF_{p}] \rightarrow \FF_{p^{r}}$. If $\Phi^{-\transp}$ is applied to a $m\times r$ matrix we assume that it operates row-wise. At this point, we make use of the following result:
\begin{lemma}\label{lem3} For $\Xm \in \FF_{p}^{m \times r}$, let $\Ym = \Qm\Xm$ for some matrix $\Qm \in \FF_{p}^{m \times m}$. Set $\yv = \Qm \xv$ where $\xv = \Phi^{-\transp}(\Xm) \in \FF_{p^r}^m$. Then, we have:
\begin{equation}
\yv = \Phi^{-\transp}(\Ym).
\end{equation}
\end{lemma}
\begin{IEEEproof} Since $\Qm$ is defined over the ground field $\FF_{p}$, we have:
\begin{equation}
\Phi^{\transp}(\Qm\xv) = \Qm\Xm=\Ym.
\end{equation} Using this fact, we can get:
\begin{eqnarray*}
\Phi^{-\transp}(\Ym) = \Phi^{-\transp}(\Phi^{\transp}(\Qm\xv) ) =\Qm\xv = \yv.
\end{eqnarray*}This completes the proof.
\end{IEEEproof}
Using Lemma~\ref{lem3}, we have:
\begin{eqnarray*}
\yv_{1} = \Phi^{-\transp}(\Ym_{1}) &=& \Qm_{11}\xv_{1} + \Qm_{12}\xv_{2}\\
&=& \Qm_{11}\Vm_{1}\wv_{1} + \Qm_{12}\Vm_{2}\wv_{2}\\
&=& \Qm_{11}\Vm_{1}\uv_{1}
\end{eqnarray*} where the last step is due to the fact that precoding vectors satisfy the alignment conditions in (\ref{eq:alignment}). Similarly, relay 2 observes the aligned signal:
\begin{eqnarray*}
\yv_{2} &=& \Qm_{21}\Vm_{1}\uv_{2}.
\end{eqnarray*} At this point, we can follow Section~\ref{sec:AND}. In this case, we can achieve the sum-rate of $(2m-1)\log{p^r}$  during $r$ time slots. Therefore, we can achieve the sum-rate of $(2m-1)\log{p}$ per time slot.

\begin{remark}The number of required symbol extensions $r\leq m!$ depends on the channel coefficients.
In general, we can always use the $m!$-symbol extension to apply AND, regardless of channel coefficients.
In this way, the coding block length (symbol extension order) depends only on the number of inputs/outputs at each node,
and it is independent of the channel coefficients.  However, this approach is only applicable for a small $m$ due to the large
symbol-extension, much larger than the minimum required symbol extension, as shown Table~\ref{table:upper}.
\hfill $\lozenge$
\end{remark}
\begin{definition}\label{def:Sg} An $m \times m$ matrix $\Qm$ over a finite field $\FF_{p}$ is said to be {\em separable} if its characteristic polynomial $c(x)$ has no repeated roots in the algebraic closure of $\FF_{p}$. Also, we define $s_{\mbox{{\tiny GL}}}(p,m)$ by the proportion of separable elements in $\mbox{GL}(p,m)$. \hfill $\lozenge$
\end{definition}
With this definition, we have:
\begin{theorem} If the channel matrices $\Qm_{\ell k}$ are independently and uniformly drawn from $\mbox{GL}(p,m)$, the sum-rate of $(2m-1)\log{p}$ is achievable with probability $\PP_{\rm{MIMO}}=(s_{\mbox{{\tiny GL}}}(p,m))^2$.
\end{theorem}
\begin{IEEEproof} The proof consists of computing the probability that feasibility conditions hold for the first hop. The same probability can be immediately applied to the second hop. Recall that the feasibility condition is equal to $\Qm=\Qm_{11}^{-1}\Qm_{12}\Qm_{22}^{-1}\Qm_{21}$ having all distinct eigenvalues. Since $\Qm_{\ell k}$ are independently and uniformly drawn from $\mbox{GL}(p,m)$, the $\Qm$ also follows the same distribution, i.e, $\Qm$ is uniformly distributed from $\mbox{GL}(p,m)$. Then, the desired probability is equal to the $s_{\mbox{{\tiny GL}}}(p,m)$ (see Definition~\ref{def:Sg}). Therefore, we have that $\PP_{\rm{MIMO}}=(s_{\mbox{{\tiny GL}}}(p,m))^2$ and this completes the proof.
\end{IEEEproof}
\begin{corollary} If each element of channel matrix is independently and uniformly drawn from $\FF_{p}$, the sum-rate of $(2m-1)\log{p}$ is achievable with probability $[\prod_{i=1}^{m}(1-p^{-i})]^{8}(s_{\mbox{{\tiny GL}}}(p,m))^2$.
\end{corollary}
\begin{IEEEproof} Each channel matrix $\Qm_{\ell k}$ is full-rank with probability $[\prod_{i=1}^{m}(1-p^{-i})]$. Under the assumption that all channel matrices are full-rank, the probability of $\Qm$ having all distinct eigenvalues is equal to $s_{\mbox{{\tiny GL}}}(p,m)$. This completes the proof.
\end{IEEEproof}

In \cite{Fulman}, Fulman derived the lower and upper bounds on $s_{\mbox{{\tiny GL}}}(p,m)$ as
\begin{equation}
1 - p^{-1} - \frac{8(p-1)}{(2p-3)}\left(\frac{2}{3}p\right)^{-m} \leq s_{\mbox{{\tiny GL}}}(p,m) \leq 1 - p^{-1} + \frac{8(p-1)}{(2p-3)}\left(\frac{2}{3}p\right)^{-m}.
\end{equation} Also, using generating function techniques, it was proved in \cite{Fulman1,Wall} that the limit exists as
\begin{equation}
s_{\mbox{{\tiny GL}}}(p,\infty) = 1- p^{-1}.\label{probf:GL}
\end{equation}

\begin{remark}\label{remark:prob} ({\em Feasibility Probability}) Using (\ref{probf:C}), (\ref{probf:D}), and (\ref{probf:GL}), we can compute the probabilities that AND is feasible for the scalar $m$-th extension field model, scalar ground-field model (with $m$-symbol extension),
and MIMO ground-field model. In the limit, i.e., either $m$ or $p$ goes to $\infty$, the feasibility probabilities are given in Table~\ref{table:FP}. We notice that only
the $m$-th extension field model satisfies the feasibility conditions with probability 1 when $p$ is finite.\hfill $\lozenge$
\end{remark}

\begin{table}
\caption{Feasibility Probabilities in the limit}
\begin{equation*}
\begin{tabular}{|c||c|c|}
\hline
   & $p \rightarrow \infty$ & $m \rightarrow \infty$ \\
   \hline
   $\PP_{\rm{FE}}$ & 1 & 1 \\
   \hline
   $\PP_{\rm{SE}}$ & 1 & 0\\
   \hline
   $\PP_{\rm{MIMO}}$ & 1 & $(1-p^{-1})^2$\\
   \hline
\end{tabular}
\end{equation*}
\label{table:FP}
\end{table}

%
\section{Two-Unicast Flows Multihop Wired Networks}\label{sec:PBNA-2}

\begin{figure}[ht]
\centerline{\includegraphics[width=16cm]{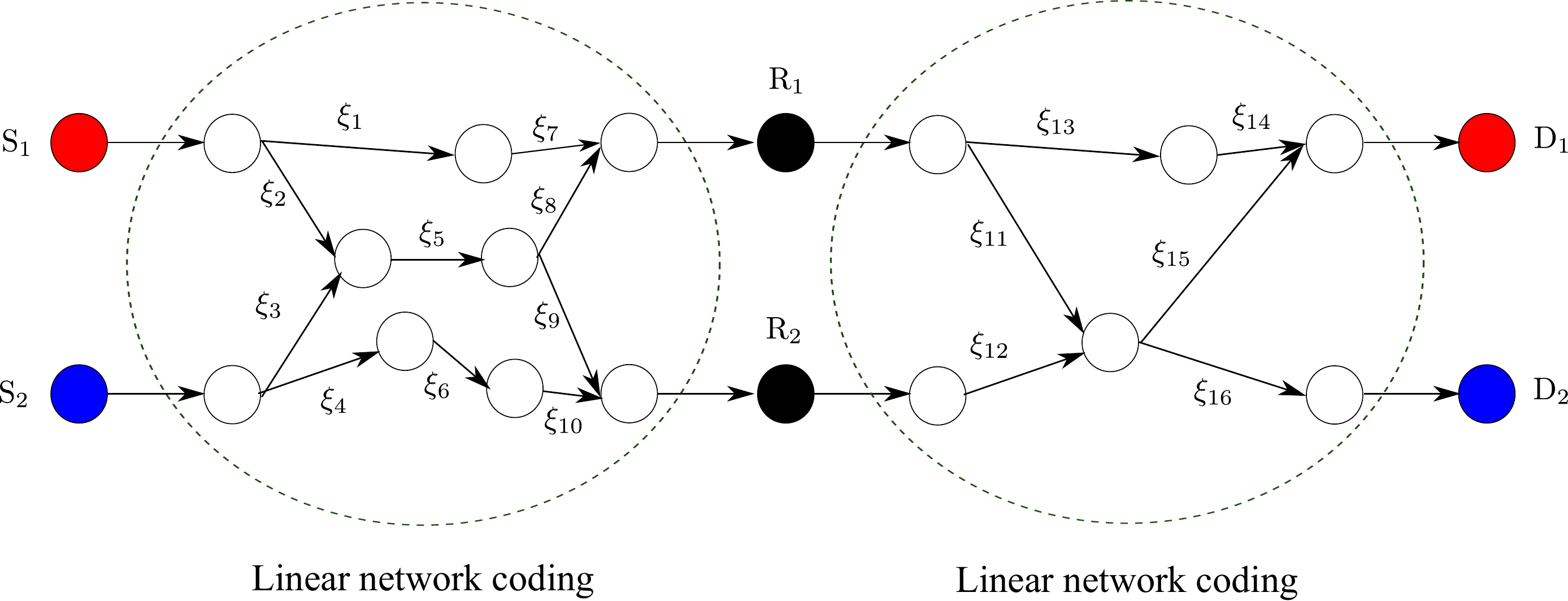}}
\caption{Wired network model as $2\times 2\times 2$ IC.}
\label{wired}
\end{figure}

We study two-flow multihop wired networks consisting of two sources, multiple intermediate relays, and two destinations. In this model,  we assume that one symbol of ground-field $\FF_{p}$ is transmitted per unit time, i.e., the capacity of a wired link is given by $\log{p}$ bits per unit time. We make use of the network topology in Fig.~\ref{wired} to explain our coding scheme.  This consists of two parts: 1) We first identity two {\em key relays} (i.e., $\mbox{R}_{1}$ and $\mbox{R}_{2}$ in Fig.~\ref{wired}) which are responsible for non-trivial relaying operations. All other relays except the key relays simply perform scalar (or vector) random linear network coding (RLNC) operations. This yields a {\em condensed} version of the network that only consists of two sources, two key relays, and two destinations. 2) AND scheme is applied to the resulting $2\times 2\times 2$ finite-field IC.

For the case of scalar RLNC, the relays combine their incoming symbols by multiplying them with scalar coefficients called {\em coding coefficients} and add them to produce new outgoing symbols. Further, in scalar RLNC, the coding coefficients are simply chosen independently and uniformly from $\FF_{p}^{\star}$ \cite{Ho}. We assume $m$ channel uses (i.e., $m$-symbol extension) for the scalar RLNC in order to use AND. Vector linear network coding was proposed in \cite{Ebrahimi} for which relays perform the coding operations over $m$-dimensional vectors, namely, multiply their incoming vectors with $m \times m$ coding matrices over $\FF_{p}$. In this paper we propose a vector RLNC where the $m \times m$ coding matrices are chosen independently and uniformly from the powers of companion matrix (i.e., as elements of $\Cc^{\star}(p,m)$). Equivalently, the operations of vector RLNC can be done over the extension-field $\FF_{p^m}$ using the vector/matrix representation of extension-field in Section~\ref{sec:MAC}. Both scalar and vector RLNC can yield the $2\times 2\times 2$ MIMO IC consisting of two sources $\mbox{S}_{1}$ and $\mbox{S}_{2}$, two intermediate relays $\mbox{R}_{1}$ and $\mbox{R}_{2}$, and two destinations $\mbox{D}_{1}$ and $\mbox{D}_{2}$. Here, the channel matrices are obtained as function of coding coefficients.

\begin{example} In the example with reference to Fig.~\ref{wired}, the channel matrices of the first hop are given by
\begin{equation}
\left[
  \begin{array}{cc}
    \Qm_{11} & \Qm_{12} \\
    \Qm_{21} & \Qm_{22} \\
  \end{array}
\right]=\left[
          \begin{array}{cc}
            \xi_{1}\xi_{7} + \xi_{2}\xi_{5}\xi_{8} & \xi_{3}\xi_{5}\xi_{8} \\
            \xi_{2}\xi_{5}\xi_{9} & \xi_{4}\xi_{10}+\xi_{3}\xi_{5}\xi_{8} \\
          \end{array}
        \right]
\end{equation} Likewise, the channel matrices of the second hop are obtained as
\begin{equation}
\left[
  \begin{array}{cc}
    \Qm_{33} & \Qm_{34} \\
    \Qm_{43} & \Qm_{44} \\
  \end{array}
\right]=\left[
          \begin{array}{cc}
            \xi_{13}\xi_{14} + \xi_{11}\xi_{15} & \xi_{12}\xi_{15} \\
            \xi_{11}\xi_{16} & \xi_{12}\xi_{16} \\
          \end{array}
        \right]
\end{equation} where notice that $\xi_{i} \in \Dc^{\star}(p,m)$ for scalar RLNC and $\xi_{i}(p,m) \in \Cc^{\star}(p,m)$ for vector RLNC. \hfill $\lozenge$
\end{example}

For the case of vector RLNC, the resulting channel matrices $\Qm_{\ell k}$ are the elements of $\Cc(p,m)$ and with high probability, they are not all-zero matrices as long as $p^m$ is large enough. However, in case of scalar RLNC, we notice that $\Qm_{\ell k}$ may contain zero diagonal elements if $p$ is not large enough. Assuming that the feasibility conditions of AND hold, scalar RLNC with AND and vector RLNC with AND can achieve $\frac{(2m-1)}{m}\log{p}$. Letting $m$ large enough, both schemes can achieve the cut-set upper bound of $2\log{p}$ subject to the feasibility probability. As seen in Table~\ref{table:FP}, this probability of vector RLNC approaches to 1 with a large $m$ irrespectively of $p$ while scalar RLNC requires to have a sufficiently large $p$ in order to have close-to-one feasibility probability. It is remarkable that vector RLNC with AND  achieves the cut-set upper bound of $2\log{p}$ for any prime $p$ (i.e., even for small capacity of wired links).

It is also interesting to classify the topologies of wired networks for which vector RLNC with AND achieves the cut-set bound of $2\log{p}$ (by letting $m$ large). Recently, all network topologies of two unicast layered wireless networks are classified in terms of degrees-of-freedom (DoF) \cite{Shomorony} and it was shown that if the channel gains are chosen independently according to continuous distributions, then, with probability 1, two-unicast {\em layered} Gaussian networks can only have 1, 3/2, or 2 sum DoFs. The coding scheme is based on linear forwarding (amplify and forward) of received signals at all relays, except for a small fraction of key relays. The linear forwarding creates {\em condensed} network that only consists of two sources, two relays, and two destinations. If the resulting condensed network is an $2\times 2\times 2$ IC,
the AND based on the framework of rational dimensions in \cite{Gou} was applied to achieve sum-DoF equal to 2.
We wish to remark that this scheme is analogous to vector RLNC with AND for the wired (deterministic) network model over finite fields. Thus, we can see that vector RLNC with AND achieves the cut-set upper bound of $2\log{p}$ for those wired networks whose
corresponding condensed network induced by the same method of \cite{Shomorony} falls into the $2\times 2\times 2$ IC equivalence class.


Focusing on more practical systems, we can consider wired networks for which the capacity of each link is equal to some value, indicated by $R_{0}$ (bits per unit time).
For fixed $R_{0}$, we can construct a scalar RLNC and a vector RLNC by choosing the ground field $\FF_{p}$ and $m$ symbol (or field) extension.
Then, the transmit signals of both schemes take on the form of $m$-dimensional vectors over the ground field $\FF_{p}$.
In order to send such vectors via the wired links, we require $\mbox{T} = \frac{m\log{p}}{R_{0}}$ units of time where $\mbox{T}$ is assumed to be an integer.
Both schemes can achieve $(2m-1)\log{p}$ bits per vector (i.e., per $\mbox{T}$ channel uses), assuming that the feasibility conditions of AND hold.
Accordingly, we have the normalized  throughput per unit time:
\begin{equation}
\mbox{Throughput} = \frac{(2m-1)\log{p}}{\mbox{T}}= \frac{(2m-1)R_{0}}{m} \mbox{ bits per unit time}.
\end{equation}
Focusing solely on high throughput, we can choose $p$ large enough such that the feasibility probability goes to 1 and also
choose $m$ large enough such that $(2m-1)/m$ is close to $2$, such that the cut-set upper bound of $2R_{0}$ is nearly achieved.
In other words, both schemes can achieve the optimal throughput by allowing a sufficiently long ``latency" $\mbox{T}=\frac{m\log{p}}{R_{0}}$.
However, notice that this latency  increases linearly with $m$ and logarithmically with $p$. When considering practical systems, the design parameters $m$ and $p$
should be carefully chosen by taking latency and throughput into account. One reasonable approach is to choose $m$ and $p$ such that both a desired
gap to the cut-set bound and a target feasibility probability are guaranteed.
From Fig.~\ref{comp}, we observe that vector RLNC allows for a smaller $p$ than scalar RLNC. For example, suppose that we wish to achieve
the cut-set bound within a gap of $\epsilon R_{0}$ and feasibility probability $1-\delta$. Then, with vector RLNC we can choose $p$ as small as $p=2$, while with scalar RLNC we should choose $p$ at least larger than $m$ (in general, much larger).
Therefore, vector RLNC can significantly reduce latency.

%
\section{Application to $2\times 2\times 2$ Scalar Gaussian IC}\label{sec:GA}

In this section we study the $2\times 2\times 2$ Gaussian IC where all nodes have a single antennas. Also, channel coefficients are time-invariant (i.e., constant over coding blocks). A block of $n$ channel uses in the first and second hops is described by
\begin{equation}
\left[
  \begin{array}{c}
   \underline{\yv}_{1} \\
   \underline{\yv}_{2} \\
  \end{array}
\right] = \left[
            \begin{array}{cc}
              h_{11} & h_{12} \\
              h_{21} & h_{22} \\
            \end{array}
          \right]\left[
  \begin{array}{c}
    \underline{\xv}_{1} \\
   \underline{\xv}_{2} \\
  \end{array}
\right] + \left[
  \begin{array}{c}
    \underline{\zv}_{1} \\
   \underline{\zv}_{2} \\
  \end{array}
\right]
\end{equation}
and by
\begin{equation}
\left[
  \begin{array}{c}
    \underline{\yv}_{3} \\
    \underline{\yv}_{4} \\
  \end{array}
\right] = \left[
            \begin{array}{cc}
              h_{33} & h_{34} \\
              h_{43} & h_{44} \\
            \end{array}
          \right]\left[
  \begin{array}{c}
   \underline{\xv}_{3} \\
    \underline{\xv}_{4} \\
  \end{array}
\right]+ \left[
  \begin{array}{c}
    \underline{\zv}_{3} \\
   \underline{\zv}_{4} \\
  \end{array}
\right],
\end{equation}
respectively,
where $\underline{\xv}_{k} \in \CC^{1\times n}$ and $\underline{\yv}_{\ell} \in \CC^{1\times n}$ denote the channel input and output vectors,
$\underline{\zv}_{\ell} \in \CC^{1 \times n}$ contains i.i.d. Gaussian noise samples $\sim \Cc\Nc(0,1)$,
and the channel coefficients $h_{\ell k}\in \CC$ are {\em constant} over the whole block of length $n$ and known to all nodes.
The same transmit power constraint  $\frac{1}{n}\mathbb{E}[\|\underline{\xv}_{k}\|^2] \leq \SNR$ is assumed for all transmitters $k=1,2,3,4$.
We use a vector representation of channel coefficients such that $\hv_{\ell} = [h_{\ell 1}, h_{\ell 2}]^{\transp}$
for $\ell=1,2$ and $\hv_{\ell}=[h_{\ell 3}, h_{\ell 4}]^{\transp}$ for $\ell=3,4$.

\begin{figure}[ht]
\centerline{\includegraphics[width=14cm]{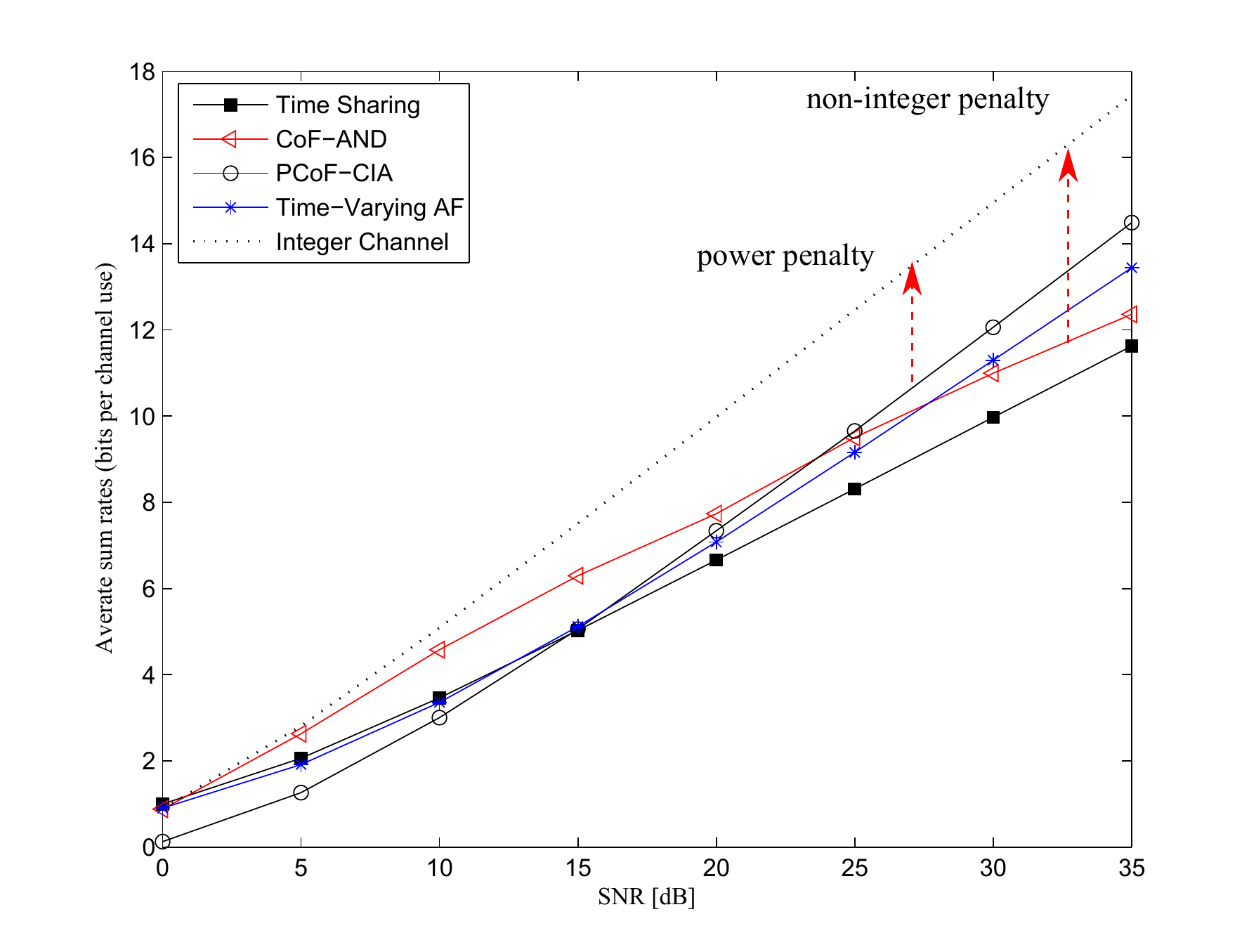}}
\caption{Achievable symmetric sum-rates of PCoF-CIA, CoF-AND, and Time Sharing for $2\times 2\times 2$ scalar Gaussian IC.}
\label{simulation}
\end{figure}

For such channels, the optimal sum-DoF equal to 2 was achieved in \cite{Gou} by using a scheme known as ``aligned interference neutralization",
based on the framework of rational dimensions \cite{Motahari}. Also, in \cite{Gou}, a linear beamforming scheme achieving sum-DoF equal to 3/2 is found, by exploiting asymmetric complex signaling \cite{Gou,Cadambe-Asy}. In this section we present two coding schemes that combine the result of Section~\ref{sec:FE} with the
compute-and-forward (CoF) framework of \cite{Nazer}, named CoF-AND (see Section~\ref{subsec:CoFwithAND}) and PCoF-CIA (see Section~\ref{subsec:PCoFwithCIA}). We show that PCoF-CIA can achieve the $3/2$ sum-DoF (same as asymmetric complex signaling in \cite{Gou}) while the achievable
sum-DoF of CoF-AND is lower than $3/2$ due to the non-integer penalty (i.e., the residual self-interference due to the fact that the channel coefficients take on non-integer values in practice and see \cite{Nazer,Hong-J} for details).  However,  in spite of  its poor DoF performance, CoF-AND may be competitive in the lower SNR range
(see Fig.~\ref{simulation}). This crossing-point can be explained as follows. In PCoF-CIA, source 1 sends two data streams while source 2 sends only one data stream.
Hence, the required transmission power at source 1 is usually higher than what required at source 2. 
Since with CoF both sources should employ
the same nested lattice code, the second moment of the lattice code is limited by the stricter transmission power constraint. 
For this reason, source 2 cannot make use of full-transmission power, which may result in a performance loss at low SNR. 
However, in the case of CoF-AND, this problem does not occur since
AND precoding is performed in the finite-field domain 
(see details in Section~\ref{subsec:CoFwithAND}). 

Fig.~\ref{simulation} shows the finite-SNR performance of PCoF-CIA and CoF-AND in terms of their achievable {\em ergodic} sum-rate, obtained by 
Monte Carlo simulation when the channel coefficients are equal to $\sqrt{\SNR} e^{j\phi_{ij}}$, where $\phi_{ij}$ are i.i.d. uniformly 
phases uniformly distributed over $[0,2\pi]$ (independent phase-fading).   For comparison, we considered the performance of {\em time-sharing} (i.e., TDMA), whose achievable sum-rate is equal to $\log(1+\SNR)$. Also, we considered the performance of the time-varying amplify-and-forward (AF) scheme proposed in 
\cite{Issa}, achieves $4/3$ sum-DoF, with finite-SNR achievable rate given in \cite[Section III]{Issa}. We observe that CoF-AND yields the best performance
for SNR below 22 dB, and  PCoF-CIA outperforms time-sharing and time-varying AF for $\SNR \geq 15$ dB, and its gain increases with SNR, reflecting the 
larger sum-DoF (i.e., the slope of the sum-rate curve at high SNR).

\subsection{Compute and Forward with Aligned Network Diagonalization}\label{subsec:CoFwithAND}

This scheme consists of two steps: 1) The CoF framework in \cite{Nazer} converts the $2\times 2\times 2$ Gaussian IC into a $2\times 2\times 2$ IC
over $\FF_{p^2}$. 2) Applying AND as in Section~\ref{sec:AND} to the resulting $2\times 2\times 2$ finite-field channel, the end-to-end interference
is eliminated in the finite-field domain. The main result of this section is given by:

\begin{theorem}\label{thm:CoFAND} CoF-AND achieves the {\em symmetric} sum-rate of $3R/2$ with all messages of the same rate given by
\begin{equation}
R = \min_{\ell=1,2,3,4}\left\{\log^{+}\left(\frac{\SNR}{\bv_{\ell}^{\herm}(\SNR^{-1}\Id+\hv_{\ell}\hv_{\ell}^{\herm})^{-1}\bv_{\ell}}\right)\right\}
\end{equation} for any full-rank integer matrices
\begin{eqnarray}
\Bm_{1}^{\herm} \in \ZZ[j]^{2\times 2} &=& \left[
            \begin{array}{cc}
              \bv_{1}^{\herm} \\
              \bv_{2}^{\herm}  \\
            \end{array}
          \right] = \left[
            \begin{array}{cc}
              b_{11} & b_{12} \\
              b_{21} & b_{22} \\
            \end{array}
          \right]\\
 \Bm_{2}^{\herm} \in \ZZ[j]^{2 \times 2} &=& \left[
            \begin{array}{cc}
              \bv_{3}^{\herm} \\
              \bv_{4}^{\herm}  \\
            \end{array}
          \right] = \left[
            \begin{array}{cc}
              b_{33} & b_{34} \\
              b_{43} & b_{44} \\
            \end{array}
          \right]
\end{eqnarray} such that $\Im(b_{11}^{-1}b_{12}b_{22}^{-1}b_{21})\neq 0$ and $\Im(b_{33}^{-1}b_{34}b_{44}^{-1}b_{43})\neq 0$.
\end{theorem}

\begin{IEEEproof} Let $\Cc$ denote a linear code over $\FF_{p^2}$ with block length $n$ and dimension $r$, with generator matrix $\Gm$. Also, we let $\Lambda$ and $\Lambda_{1}$ denote nested lattices constructed as
\begin{equation}
\Lambda_{1} = p^{-1}g(\Cc)\Tm + \Lambda
\end{equation} where  $\Lambda=\{\underline{\lambda}=\underline{\zv}\Tm: \underline{\zv} \in \ZZ^{n}[j]\}$ with full-rank generator matrix $\Tm \in \CC^{n \times n}$ and $g(\Cc)$ is the image of $\Cc$ under the mapping $g$ (applied component-wise).  All users make use of the same nested lattice codebook $\Lc=\Lambda_{1} \cap \Vc_{\Lambda}$ where $\Lambda$ has second moment $\sigma_{\Lambda}^{2}=\SNR$.
The natural labeling function $f:\FF_{p^2}^{r} \rightarrow \Lc$ is defined by $f(\underline{\wv}_{k}) = [p^{-1}g(\underline{\wv}\Gm)\Tm] \mod \Lambda$
and maps information messages $\wv \in \FF_{p^2}^r$ onto lattice codewords (vectors in $\Lc$).
The lattice coding rate is given by $R = \frac{2r}{n}\log p$.
Following the CoF framework (see \cite{Nazer} for details), each relay $\ell$ can reliably decode\footnote{We use the term ``reliably decoding'' for brevity, in order to indicate that there exist sequences of lattice codes for increasing $n$ for which the probability of decoding error vanishes as $n \rightarrow \infty$, if the rate conditions (\ref{cof-rate}), (\ref{cof-rate1}) are satisfied.}
a linear combination $\underline{\vv}_{\ell} = [\sum_{k=1}^{2}b_{\ell k} f(\underline{\wv}_{k})] \mod \Lambda$ with integer coefficient vector $\bv_{\ell} = [b_{\ell 1}, b_{\ell 2}]$, provided that
\begin{equation}  \label{cof-rate}
R \leq \log^{+}\left(\frac{\SNR}{\bv_{\ell}^{\herm}(\SNR^{-1}\Id + \hv_{\ell}\hv_{\ell}^{\herm})^{-1}\bv_{\ell}}\right) \mbox{ for } \ell = 1,2.
\end{equation}
Using the linearity of lattice encoding, the corresponding linear combinations over $\FF_{p^2}$ are given by $\underline{\uv}_{{\rm R}_{\ell}} = \Qc_{\ell 1}\underline{\wv}_{1} + \Qc_{\ell 2} \underline{\wv}_{2}$ where $\Qc_{\ell k} = g^{-1}([b_{\ell k}] \mod p\ZZ[j])$ for $k,\ell=1,2$. Relay $\ell$ forwards the decoded linear combination $\underline{\uv}_{\ell}$ using the same lattice code.  Likewise, destination $\ell$ can reliably decode a linear combination of relays' messages $\underline{\uv}_{{\rm D}_{\ell}} =\Qc_{\ell 3}\underline{\uv}_{{\rm R}_{1}} + \Qc_{\ell 4}\underline{\uv}_{{\rm R}_{2}}$ if
\begin{equation} \label{cof-rate1}
R \leq   \log^{+}\left(\frac{\SNR}{\bv_{\ell}^{\herm}(\SNR^{-1}\Id + \hv_{\ell}\hv_{\ell}^{\herm})^{-1}\bv_{\ell}}\right) \mbox{ for } \ell = 3,4
\end{equation}
where  $\Qc_{\ell k} = g^{-1}([b_{\ell k}] \mod p\ZZ[j])$ for $k,\ell=3,4$. Based on this, we can define a {\em noiseless} $2\times 2\times 2$ IC over $\FF_{p^2}$  with channel coefficients $\Qc_{\ell k}$, equivalent to the scalar $2$-nd extension field model in (\ref{model:1-hop}) and (\ref{model:2-hop}). Then, AND can be applied to the resulting finite-field model. From Theorem~\ref{thm:222-Field}, we can see that  this scheme achieves sum-rate of $3R/2$ as long as the feasibility conditions of AND in Theorem~\ref{thm:222-Field} hold. These feasibility conditions are addressed by the following result:

\begin{lemma}\label{lem11}
If $\Im(b_{11}^{-1}b_{12}b_{22}^{-1}b_{21}) \neq 0$ then the AND  feasibility conditions of Theorem~\ref{thm:222-Field} are satisfied.
\end{lemma}
\begin{IEEEproof} Recall that $\gamma = \Qc_{11}^{-1}\Qc_{12}\Qc_{22}^{-1}\Qc_{21}$ where $\Qc_{\ell k} = g^{-1}([b_{\ell k}] \mod p\ZZ[j]) \in \FF_{p^2}$. In order to guarantee the feasibility condition in Theorem \ref{thm:222-Field}, we need to show that $\gamma$ has degree-2 minimal polynomial.
Writing $b_{11}^{-1}b_{12}b_{22}^{-1}b_{21} = \frac{b_{11}^*b_{12}b_{22}^*b_{21}}{|b_{11}|^2|b_{22}|^2}$, where $^*$ denotes complex conjugate,
we notice that  $|b_{11}|^2$ and $|b_{22}|^2$ are positive integer since $b_{11},b_{22}$ are integers.
Hence, we can always choose $p_{1},p_{2} \in \ZZ_{+}$
such that $c_{1} =|b_{11}|^2p_{1}$, $c_{2}=|b_{22}|^2p_{2}$ and
\begin{equation}
[c_{1}] \mod p\ZZ = 1\label{eq:1}
\end{equation}
\begin{equation}
[c_{2}] \mod p\ZZ = 1.\label{eq:2}
\end{equation}
Define $b_{11}^{\dag} = [c_{1}b_{11}^{-1}] \mod p\ZZ[j]$ and $b_{22}^{\dag} = [c_{2}b_{22}^{-1}] \mod p\ZZ[j]$. Using (\ref{eq:1}) and (\ref{eq:2}), we have that
\begin{equation}
\Qc_{kk}^{-1} = g^{-1}([b_{kk}^{\dag}] \mod p\ZZ[j]), k=1,2\label{eq:invq}
\end{equation} since
\begin{eqnarray*}
\Qc_{kk}\Qc_{kk}^{-1} &=& g^{-1}([b_{kk}] \mod p\ZZ[j])g^{-1}([b_{kk}^{\dag}] \mod p\ZZ[j]) \\
&=& g^{-1}([b_{kk}b_{kk}^{\dag}] \mod p\ZZ[j])\\
&=& g^{-1}([c_{k}] \mod p\ZZ[j]) = 1.
\end{eqnarray*} Using (\ref{eq:invq}), we have:
\begin{eqnarray*}
g(\gamma) &=&  [b_{11}^{\dag}b_{12}b_{22}^{\dag}b_{21}] \mod p\ZZ[j]\\
&=&[c_{1}b_{11}^{-1}b_{12}c_{2}b_{22}^{-1}b_{21}] \mod p\ZZ[j] \\
&=& [p_{1}p_{2} (b_{11}^{*}b_{12}b_{22}^{*}b_{21})] \mod p\ZZ[j].
\end{eqnarray*} Because of $p_{1}, p_{2} \in \ZZ_{+}$, we have that $\Im(g(\gamma))=[p_{1}p_{2}\Im (b_{11}^{*}b_{12}b_{22}^{*}b_{21})] \mod p\ZZ $. Since $p$ is a prime, $p_{1}$ and $p_{2}$ are not zero and by assumption $\Im (b_{11}^{*}b_{12}b_{22}^{*}b_{21}) \neq 0$, then $\Im(g(\gamma)) = 0$ if and only if at least one term $p_{1},p_{2}$ or $\Im (b_{11}^{*}b_{12}b_{22}^{*}b_{21})$ is a multiple of $p$. The coefficients $p_{k}$ are not multiple of $p$ since $[|b_{kk}|^2p_{k}] \mod p\ZZ = 1$ for $k=1,2$. Also, $\Im (b_{11}^{*}b_{12}b_{22}^{*}b_{21})$ is not a multiple of $p$ for $p$ sufficiently large.\footnote{Notice that in the CoF construction the prime $p$ in the sequence of lattice codes that achieves the computation rates (\ref{cof-rate}), (\ref{cof-rate1}) increases without bounds as $n \rightarrow \infty$ (see \cite{Nazer} and references therein).}
Therefore, we conclude that $\Im(g(\gamma)) \neq 0$. It follows that $\gamma$ cannot be an element of $\FF_{p} = \ZZ_{p}$, and therefore its minimal polynomial has degree 2.
\end{IEEEproof}

The same condition applies to $\gamma' = \Qc_{33}^{-1}\Qc_{34}\Qc_{44}^{-1}\Qc_{43}$ for the second hop where and $\Qc_{\ell k} = g^{-1}([b_{\ell k}] \mod p\ZZ[j]) \in \FF_{p^2}$, and this concludes the proof of Theorem~\ref{thm:CoFAND}.

\end{IEEEproof}

\subsection{Precoded Compute and Forward with Channel Integer Alignment}\label{subsec:PCoFwithCIA}

\begin{figure}[ht]
\centerline{\includegraphics[width=14cm]{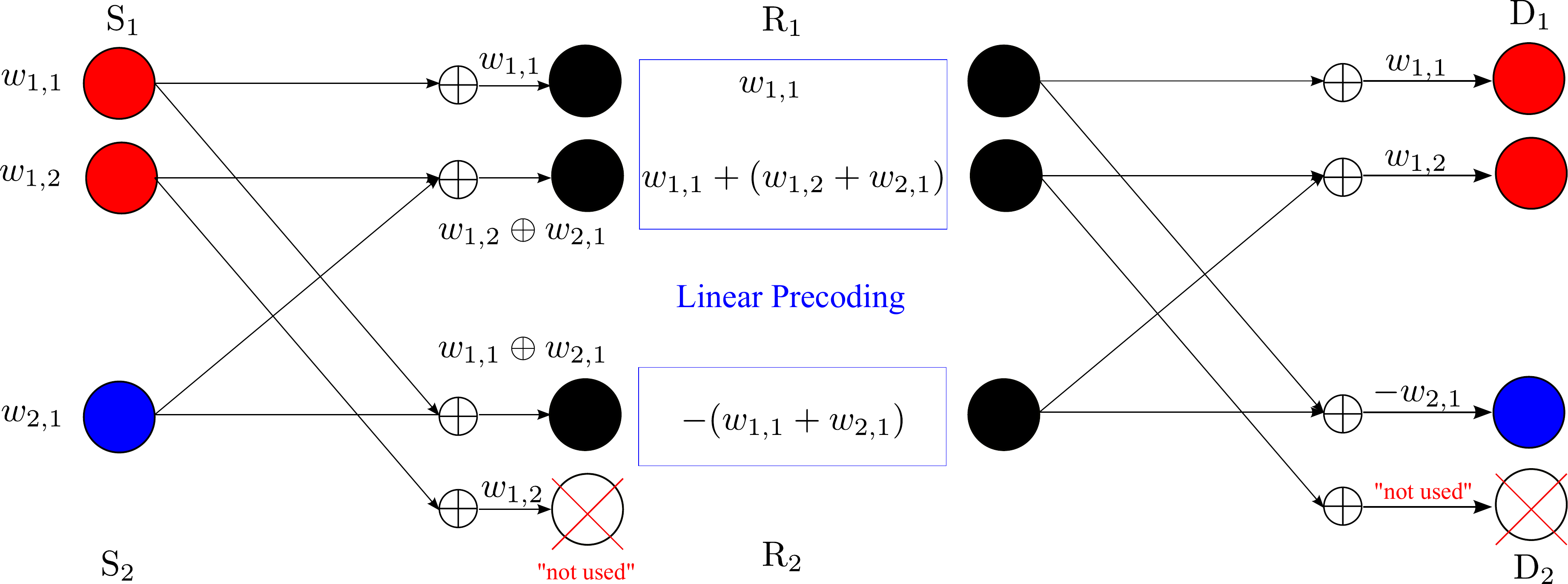}}
\caption{A deterministic $2\times 2\times 2$ finite field interference channel.}
\label{precoding}
\end{figure}

The main idea of this scheme is to see the complex field $\CC$ as the extension-field with ground field $\RR$. Namely, for any $\Xc \in \CC$, we can define its vector and matrix representations:
\begin{eqnarray}
\Phi(\Xc) &=& [\Re(\Xc),\Im(\Xc)]^{\transp} \in \RR^2\\
\Gamma(\Xc) &=& \left[
                 \begin{array}{cc}
                   \Re(\Xc) & -\Im(\Xc)  \\
                   \Im(\Xc)  & \Re(\Xc) \\
                 \end{array}
               \right].
\end{eqnarray}
Using the above mappings, we can yield the $2\times 2\times 2$ MIMO IC over $\RR$ where all nodes have 2 multiple antennas. The concept of using complex signal as $2$-dimensional real signal was also used in \cite{Cadambe-Asy} and it is known as ``asymmetric complex signaling". In the context of the problem at hand,
we use our recently proposed PCoF-CIA scheme for the $2\times 2\times 2$ MIMO IC \cite{Hong-J}.
This scheme can be applied to the resulting MIMO IC over $\RR$ with channel matrices
\begin{equation}
\Gamma(h_{\ell k}) \eqdef \left[
                 \begin{array}{cc}
                   \Re(h_{\ell k}) & -\Im(h_{\ell k})  \\
                   \Im(h_{\ell k})  & \Re(h_{\ell k}) \\
                 \end{array}
               \right] \in \RR^{2\times 2}.
\end{equation} Notice that for transmission, we need to transform the $2$-dimensional real-valued vector into the scalar complex-value using the inverse function of $\Phi$. PCoF-CIA can be summarized as follows (see \cite{Hong-J} for more details):
\begin{itemize}
\item  The main role of CIA (i.e., precoding over $\RR$) is to convert a real valued channel into a integer-valued channel. Namely, we choose the precoding matrices $\Vm_{1}$ and $\Vm_{2}$ at the two source nodes such that
    \begin{equation}
    \left[
      \begin{array}{cc}
        \Gamma(h_{k1})\Vm_{1} & \Gamma(h_{k2})\Vm_{2} \\
      \end{array}
    \right] = \Hm_{k}\Cm_{k}
    \end{equation} where $\Hm \in \RR^{2 \times 2}$ and $\Cm_{k} \in \ZZ^{2\times 4}$.  Using the Integer Forcing Receiver approach
    in \cite{Zhan}, each receiver $k$ can decode two linear combinations of lattice codewords with integer coefficients $\Cm_{k}$.
    This transforms {\em noisy} integer-valued IC into a {\em deterministic} finite-field IC as shown in Fig.~\ref{precoding}.
\item A linear precoding scheme is used over the finite field domain in order to eliminate the end-to-end interference (see Fig.~\ref{precoding}).
\end{itemize}
Linear precoding over $\RR$ may produce a power penalty due to the non-unitary nature of the precoding matrices of CIA, and this can degrade the performance at finite SNR. In order to counter this effect, we use Integer Forcing Beamforming  (see \cite{Hong-DAS}).
The main idea is that $\Vm_{k}$ can be pre-multiplied (from the left) by some appropriately chosen full-rank integer matrix $\Am_{k}$ since its effect can be undone by precoding over $\FF_{q}$. Then, we can optimize the integer matrix in order to minimize the power penalty of CIA. We have:

\begin{theorem}\label{thm:PCoF} PCoF-CIA can achieve the {\em symmetric} sum-rate of $3R$ with all messages of the same rate given by
\begin{eqnarray*}
R &=& \min_{\ell=1,2}\left\{\min_{k=1,2} \left\{\frac{1}{2}\log^{+}\left(\frac{\SNR}{\trace(\Vm_{1}\Am_{1}\Am_{1}^{\herm}\Vm_{1}^{\herm})\sigma_{\rm{eff},k,\ell}^2}\right)\right\},\right.\\
&&\left.\min_{k=3,4}\left\{\frac{1}{2}\log^{+}\left(\frac{\SNR}{\trace(\Vm_{3}\Am_{2}\Am_{2}^{\herm}\Vm_{3}^{\herm})\sigma_{\rm{eff},k,\ell}^2}\right)\right\}\right\}
\end{eqnarray*} for any full-rank integer matrices $\Am_{1},\Am_{2} \in \ZZ^{2 \times 2}$ and $\Bm_{k}\in \ZZ^{2 \times 2}$ with columns $\bv_{k,\ell}$, for $k=1,2,3,4$, where
\begin{equation*}
\sigma_{\rm{eff},k,\ell}^2 = \bv_{k,\ell}^{\herm}\Cm_{k}(\SNR^{-1}\Id + \Cm_{k}^{\herm}\Hm_{k}^{\herm}\Hm_{k}\Cm_{k})^{-1}\Cm_{k}^{\herm}\bv_{k,\ell}
\end{equation*}
\begin{eqnarray*}
\Hm_{k} &=& \Gamma(h_{k 1})\Vm_{1}, \Cm_{k}=\left[
                                       \begin{array}{cc}
                                         \Am_{1} & \Cm_{k2} \\
                                       \end{array}
                                     \right], k=1,2\\
\Hm_{k} &=& \Gamma(h_{k 3})\Vm_{3}, \Cm_{k} = \left[
                                       \begin{array}{cc}
                                         \Am_{2} & \Cm_{(k-2)2} \\
                                       \end{array}
                                     \right], k=3,4\\
\Vm_{1}&=& \left[
                                       \begin{array}{cc}
                                        \Gamma(h_{21}^{-1}h_{22})\frac{\onev}{\sqrt{2}} &  \Gamma(h_{11}^{-1}h_{12})\frac{\onev}{\sqrt{2}}  \\
                                       \end{array}
                                     \right],\\
\Vm_{3} &=& \left[
                                       \begin{array}{cc}
                                        \Gamma(h_{43}^{-1}h_{44})\frac{\onev}{\sqrt{2}}  &  \Gamma(h_{33}^{-1}h_{34})\frac{\onev}{\sqrt{2}}  \\
                                       \end{array}
                                     \right]\\
\Cm_{12} &\eqdef& \left[
            \begin{array}{cc}
             0  & 1\\
            \end{array}
          \right]^{\transp},\Cm_{22} \eqdef \left[
            \begin{array}{cc}
             1 & 1\\
            \end{array}
          \right]^{\transp}.
\end{eqnarray*}  \hfill\QED
\end{theorem}

\begin{corollary} PCoF-CIA achieves sum-DoF equal to $3/2$ for the $2\times 2\times 2$ scalar IC with constant channel coefficients.
\end{corollary}
\begin{IEEEproof} The proof consists of showing that $R$ defined in Theorem~\ref{thm:PCoF} grows as $\frac{1}{2}\log{\SNR}$ when $\SNR \rightarrow \infty$. In the proof of \cite[Cor. 1]{Hong-J}, we have shown that $R$ grows as $\log{\SNR}$ when $\SNR \rightarrow \infty$ as long as  $\Vm_{1}$ and $\Vm_{3}$ have full rank.  Since we use the same coding scheme here, we only need to show that $\Vm_{1}$ and $\Vm_{3}$ defined in Theorem~\ref{thm:PCoF} have full-rank.  This proof is different from what we have done in \cite[Cor. 1]{Hong-J}  since channel matrices have different structures.                                                                                                                                                                                                                                                                                                                                          
In order to show that $\Vm_{1}$ and $\Vm_{3}$ have full rank, we write:
\begin{eqnarray}
\det(\Vm_{1}) &=& \frac{1}{2} \det(\Gamma(h_{21}h_{22}^{-1}))\det\left(\left[
                                                      \begin{array}{cc}
                                                        \onev & \Gamma(h_{11}^{-1}h_{12}h_{21}h_{22}^{-1})\onev \\
                                                      \end{array}
                                                    \right]\right)\\
&=&\det(\Gamma(h_{21}h_{22}^{-1}))(\Im(h_{11}^{-1}h_{12}h_{21}h_{22}^{-1})).
\end{eqnarray} Then, we have that $\det(\Gamma(h_{21}h_{22}^{-1})) \neq 0$ and $\Im(h_{11}^{-1}h_{12}h_{21}h_{22}^{-1}) \neq 0$ with probability 1 (for continuously distributed channel coefficients), and hence $\det(\Vm_{1}) \neq 0$. Likewise, we can show that $\det(\Vm_{3}) \neq 0$. This completes the proof.
\end{IEEEproof}

%
\section{Three User Finite-Field Interference Channel}\label{sec:3-User}

\begin{figure}[ht]
\centerline{\includegraphics[width=6cm]{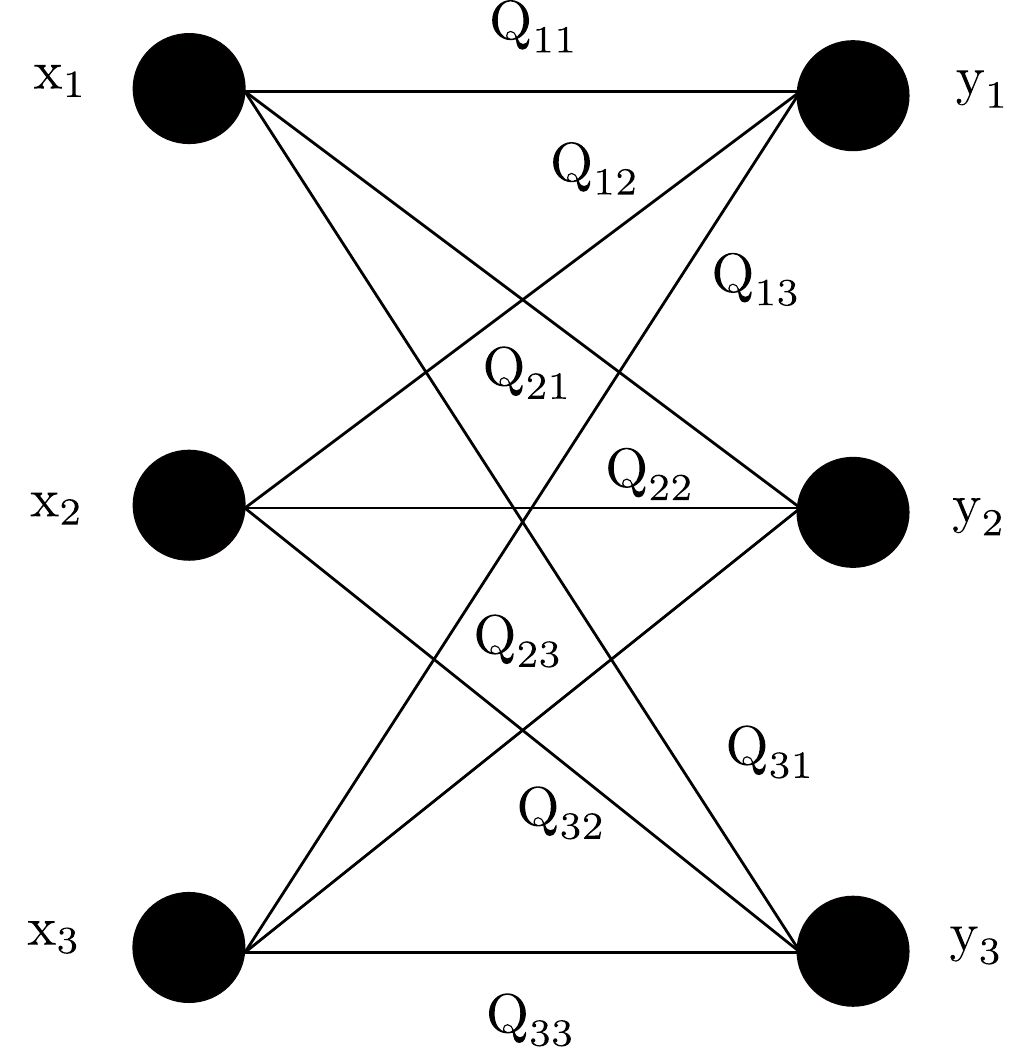}}
\caption{Three user interference channel over finite-field.}
\label{3user}
\end{figure}

For the Gaussian 3-user MIMO IC, an eigenvector-based IA scheme achieving optimal sum DoFs with probability 1 (assuming continuously distributed channel coefficients) was proposed in \cite{Cadambe08} by Cadambe and Jafar. In this section we examine the feasibility of such scheme for the finite-field counterparts as illustrated in Fig.~\ref{3user}. We start by reviewing the eigenvector-based IA scheme assuming that $m$ is even (see \cite{Cadambe08}), stated directly in the finite-field domain. We
formulate the scheme in general terms, considering $m$-dimensional input/output symbols,
such that the scheme applies to the three finite-field models considered in this paper (see Section \ref{sec:SM}).
Let $\Vm_{k} \in \FF_{p}^{m \times \frac{m}{2}}$ denote the precoding matrix used at source $k$. Then, the received signal at destination $\ell$ can be written as
\begin{equation}
\yv_{\ell} = \Qm_{\ell 1}\Vm_{1}\xv_{1} + \Qm_{\ell 2}\Vm_{2}\xv_{2} + \Qm_{\ell 3}\Vm_{3}\xv_{3}
\end{equation} where $\Qm_{k \ell} \in \FF_{p}^{m \times m}$, $\xv_{k} \in \FF_{p}^{\frac{m}{2}}$, and $\yv_{\ell} \in \FF_{p}^{m}$.
The following three conditions ensure that all interferences are aligned at destinations and the dimension of the interference is equal to the $m/2$ at all destinations:
\begin{eqnarray}
\mbox{span}(\Qm_{12}\Vm_{2}) &=& \mbox{span}(\Qm_{13}\Vm_{3})\\
\Qm_{21}\Vm_{1} &=& \Qm_{23}\Vm_{3}\\
\Qm_{31}\Vm_{1} &=& \Qm_{32}\Vm_{2}
\end{eqnarray} where $\mbox{span}(\Am)$ denotes the vector space spanned by the columns of matrix $\Am$. Also, the above conditions can be equivalently represented as
\begin{eqnarray}
\mbox{span}(\Vm_{1}) &=& \mbox{span}(\Qm\Vm_{1})\label{3-user:IA1}\\
\Vm_{2} &=& \Qm_{32}^{-1}\Qm_{31}\Vm_{1}\label{3-user:IA2}\\
\Vm_{3} &=& \Qm_{23}^{-1}\Qm_{21}\Vm_{1}\label{3-user:IA3}
\end{eqnarray} where $\Qm =\Qm_{31}^{-1}\Qm_{32}\Qm_{12}^{-1}\Qm_{13}\Qm_{23}^{-1}\Qm_{21}$.
In the eigenvector-based IA scheme \cite{Cadambe08}, the precoding matrix $\Vm_{1}$ is chosen as
\begin{equation}\label{eq:V11}
\Vm_{1} = \left[
            \begin{array}{ccc}
              \ev_{1} & \cdots & \ev_{\frac{m}{2}} \\
            \end{array}
          \right]
\end{equation} where $\ev_{1},\ldots,\ev_{\frac{m}{2}}$ denote $m/2$ linearly independent eigenvectors of $\Qm$. This choice satisfies the alignment conditions in~(\ref{3-user:IA1})-(\ref{3-user:IA3}). As a consequence, the other precoding matrices $\Vm_{2}$ and $\Vm_{3}$ are determined by~(\ref{3-user:IA2}) and~(\ref{3-user:IA3}), respectively. In order to resolve the desired signals at  the destinations, the {\em rank conditions} $\mbox{Rank}(\Sm_{k}) = m$ for $k=1,2,3$ must hold,
where
\begin{eqnarray}
\Sm_{1}=\left[
            \begin{array}{cc}
              \Qm_{11}\Vm_{1} & \Qm_{13}\Vm_{3}\\
            \end{array}
          \right]  &=& \left[
            \begin{array}{cc}
             \Vm_{1} & \Qm_{1}\Vm_{1}\\
            \end{array}
          \right]\nonumber\\
\Sm_{2}=\left[
            \begin{array}{cc}
              \Qm_{22}\Vm_{2} & \Qm_{21}\Vm_{1}\\
            \end{array}
          \right]  &=& \left[
            \begin{array}{cc}
              \Qm_{2}\Vm_{1} & \Vm_{1}\\
            \end{array}
          \right]\nonumber\\
\Sm_{3}=\left[
            \begin{array}{cc}
              \Qm_{33}\Vm_{3} & \Qm_{31}\Vm_{1}\\
            \end{array}
          \right]  &=& \left[
            \begin{array}{cc}
              \Qm_{3}\Vm_{1} &\Vm_{1}\\
            \end{array}
          \right]\label{eq:mat}
\end{eqnarray}
and where $\Qm_{1}=\Qm_{11}^{-1}\Qm_{13}\Qm_{23}^{-1}\Qm_{21}$, $\Qm_{2}=\Qm_{21}^{-1}\Qm_{22}\Qm_{32}^{-1}\Qm_{31}$, and $\Qm_{3}=\Qm_{31}^{-1}\Qm_{33}\Qm_{23}^{-1}\Qm_{21}$.

In the following sections, the feasibility of this eigenvector-based IA scheme is studied for the three different models of finite-field channel at hand.
First, we show that this scheme is feasible for the MIMO ground-field model with an appropriate symbol extension (i.e., by coding over multiple time slots). 
In case of the scalar ground-field model with $m$-symbol extension, we show that this scheme is not feasible since any eigenvector of $\Qm$ is also 
an eigenvector of $\Qm_{k}$ for $k=1,2,3$ and the rank conditions do not hold. 
Finally we provide also a negative result for the scalar $m$-th extension field model due to the same reason. 
Independently from our work, a slightly weaker analogous negative result has been shown in \cite{Krishnamurthy}, where the authors do not allow to find 
the eigenvalues of $\Qm$ in an extension field, i.e., they impose eigenvalues only in the ground field. Thus, they are restricted to consider 
only the case of  $\Gamma^{-1}(\Qm) \in \FF_{p}$, i.e., where $\Qm$ is a scalar matrix (i.e., a scaled identity matrix). 
In our work, we provide a stronger result showing that the eigenvector-based IA scheme is not feasible although we can always find eigenvalues 
of $\Qm$ in some extension field.

\subsection{MIMO ground-field model}\label{subsec:3user-MIMO}

For the time being, we assume that $\Qm$ has $m$ distinct eigenvalues. Let $c(x)$ denote the characteristic polynomial of $\Qm$. This is a degree-$m$ polynomial over $\FF_{p}$. If $c(x)$ is irreducible over $\FF_{p}$ then we can find $m$ distinct roots of $c(x)$ (i.e., $m$ distinct eigenvalues) in the extension-field $\FF_{p^m}$, given by $\{\alpha,\alpha^p,\ldots,\alpha^{p^{m-1}}\}$ where $\alpha$ is a primitive element of $\FF_{p^m}$. As explained in Section~\ref{sec:MIMO}, in general, we are able to find $m$ roots of $c(x)$ in the splitting field $\FF_{p^r}$ where the $r$ depends on channel coefficients (see Table~\ref{table:upper} for the upper bounds on $r$).
Let $\ev_{i} \in \FF_{p^r}^{m}$ denote the $i$-th eigenvector of $\Qm$. Then, we choose
\begin{equation}
\Vm_{1} = \left[
            \begin{array}{ccc}
              \ev_{1} & \cdots & \ev_{\frac{m}{2}} \\
            \end{array}
          \right] \in \FF_{p^r}^{m \times \frac{m}{2}}.
\end{equation} Notice that the alignment conditions in~(\ref{3-user:IA1})-(\ref{3-user:IA3}) are satisfied in the extension field $\FF_{p^r}$. Also the other precoding matrices are determined by
\begin{eqnarray}
\Vm_{2} &=& \Qm_{32}^{-1}\Qm_{31}\Vm_{1} \in \FF_{p^r}^{m \times \frac{m}{2}}\\
\Vm_{3} &=& \Qm_{23}^{-1}\Qm_{21}\Vm_{1}\in \FF_{p^r}^{m \times \frac{m}{2}}.
\end{eqnarray}

Assuming that $\Sm_{\ell}$ are full-rank over $\FF_{p^r}$, we present the coding scheme over the $r$-symbol extension (i.e., over $r$ time slots). Let $\xv_{k}[t]$ denote the input vector of the source $k$ at the $t$-th time slot.

\subsubsection{Encoding at the sources}

Each source $k$ precodes its message $\wv_{k} \in \FF_{p^r}^{\frac{m}{2}}$ using precoding matrix $\Vm_{k} \in \FF_{p^r}^{m\times \frac{m}{2}}$, such that
\begin{equation}
\Xc_{k} = \Vm_{k}\wv_{k} \in \FF_{p^r}^{m}
\end{equation} and transmits the $t$-th column of $[\xv_{k}[1],\ldots, \xv_{k}[r]]=\Phi^{\transp}(\Xc_{k}) \in \FF_{p}^{m\times r}$ at time slot $t$ for $t=1,\ldots,r$, where $\Phi^{\transp}: \FF_{p^r} \rightarrow [\FF_{p},\ldots,\FF_{p}]$.

\subsubsection{Decoding at the destinations}

Each destination $\ell$ observes
\begin{eqnarray*}
[\yv_{\ell}[1],\ldots, \yv_{\ell}[r]] &=& \Qm_{\ell 1}[\xv_{1}[1],\ldots, \xv_{1}[r]] + \Qm_{\ell 2}[\xv_{2}[1],\ldots,\xv_{2}[r]]\\
&& + \Qm_{\ell 3}[\xv_{3}[1],\ldots, \xv_{3}[r]].
\end{eqnarray*} From Lemma~\ref{lem3}, the received signal is mapped onto the element of $\FF_{p^r}$:
\begin{eqnarray*}
\Yc_{1} &=& \Qm_{11}\Xc_{1} + \Qm_{12}\Xc_{2}+\Qm_{13}\Xc_{3}\\
&=& \Qm_{11}\Vm_{1}\wv_{1} + \Qm_{13}\Vm_{3}(\wv_{2} + \wv_{3})\\
\Yc_{2}&=&\Qm_{22}\Vm_{2}\wv_{2} + \Qm_{21}\Vm_{1}(\wv_{1}+\wv_{3})\\
\Yc_{3}&=&\Qm_{33}\Vm_{3}\wv_{3} + \Qm_{31}\Vm_{1}(\wv_{1} + \wv_{2}).
\end{eqnarray*} Since $\Sm_{\ell}$ is assumed to be full-rank, each destination $\ell$ can decode its desired message using the matrix inversion of $\Sm_{\ell}$.

Based on this construction, we have:
\begin{theorem} If the channel matrices $\Qm_{\ell k}$ are independently and uniformly drawn from $\mbox{GL}(p,m)$, the normalized sum-capacity
\begin{equation}
d_{\rm{sum}} = \lim_{p \rightarrow \infty}\frac{C_{\rm{sum}}}{m\log{p}} = \frac{3}{2}
\end{equation} is achievable with probability 1.
\end{theorem}
\begin{IEEEproof} The upper bound is provided in \cite{Cadambe08} and the achievable scheme is given above, using the eigenvector based IA scheme with proper symbol-extension. We only need to prove that $\Sm_{\ell}$ has full-rank over $\FF_{p^r}$ with probability growing to 1 as $p \rightarrow \infty$. Following the same argument of \cite{Cadambe08}, this holds because $\Qm_{k}$ is a random (full-rank) linear transformation. The proof for $m$ odd follows along the same lines and it is therefore omitted.
\end{IEEEproof}

\subsection{Scalar ground-field model}

We consider the $m$-symbol extension with time-varying coefficients and show that the eigenvector-based IA scheme is not feasible with high probability. Suppose that channel coefficients are independently and uniformly drawn from $\FF_{p}^{\star}$. Then the channel matrices obtained by symbol-extension are diagonal matrices with all distinct diagonal elements with high probability for sufficiently large $p$. We notice that such diagonal matrices have the same eigenvectors $\ev_{1},\ldots,\ev_{m}$ such that $\ev_{i}$ denotes the all-zero vector with a single $1$ in position $i$. Hence, any eigenvector of $\Qm$ is also an eigenvector of $\Qm_{k}$ for $k=1,2,3$ and the rank conditions do not hold. In the special case of a symmetric interference channel where all interfering links have the same coefficient, the eigenvector-based IA can be applied. In fact, in this case, $\Qm$ is a scaled identity matrix and $\Qm_{k}$ has all distinct diagonal elements with high probability.
Therefore, we can always find $\frac{m}{2}$ linearly independent eigenvectors of $\Qm$ that are not the eigenvectors of $\Qm_{k}$ since any vector
is an eigenvector of a scaled identity matrix.

\subsection{Scalar $m$-th extension field model}\label{subsec:3user-field}

In this model, the channel matrices $\Qm_{\ell k}$ are represented by the powers of companion matrix $\Cm=\Gamma(\alpha)$ where $\alpha$ denotes a primitive element of $\FF_{p^m}$. That is, we have:
\begin{eqnarray}
\Qm &=& \Cm^{\ell}\\
\Qm_{k} &=& \Cm^{\ell_{k}}
\end{eqnarray} for some $\ell, \ell_{k}$, $k=1,2,3$. By considering the simple case of $m=2$, we have:
\begin{theorem}\label{thm3} For the 3-user scalar IC over $\FF_{p^2}$, we have:
\begin{description}
\item[(a)] If $\Gamma^{-1}(\Qm) \in \FF_{p}$ and $\Gamma^{-1}(\Qm_{k}) \notin \FF_{p}$ for $k=1,2,3$, the eigenvector-based IA scheme is feasible
\item[(b)] If $\Gamma^{-1}(\Qm) \notin \FF_{p}$, the eigenvector-based IA scheme is not feasible.
\end{description}
\end{theorem}
\begin{IEEEproof} From \cite{Mac}, we know that the characteristic polynomial of $\Cm$ coincides with the primitive polynomial that generates the extension-field $\FF_{p^2}$. Since the primitive polynomial is irreducible over the ground field $\FF_{p}$, we can find two distinct its roots $\lambda_{1}$ and $\lambda_{2}$ in the extension field $\FF_{p^2}$. In fact, we have that $\lambda_{1} = \alpha$ and $\lambda_{2} = \alpha^{p}$. Also, we know that the companion matrix $\Cm$ with distinct eigenvalues is diagonalizable as
\begin{equation}
\Am\Cm\Am^{-1}=\left[
                 \begin{array}{cc}
                   \alpha & 0 \\
                   0 & \alpha^p \\
                 \end{array}
               \right]=\diag(\alpha,\alpha^p)
\end{equation} where
\begin{equation}
\Am = \left[
        \begin{array}{cc}
          1 & \alpha \\
          1 & \alpha^p \\
        \end{array}
      \right]
\end{equation} is a Vandermonde matrix. Hence, for nay $\ell$, we have:
\begin{equation}\label{eq:Cm}
\Cm^{\ell} = \Am^{-1}\diag(\alpha^{\ell},\alpha^{p\ell})\Am.
\end{equation} We observe that any eigenvector of $\Cm$ is also an eigenvector of $\Cm^{\ell}$ for any $\ell$. If $\alpha^{\ell} \neq \alpha^{p\ell}$ (over $\FF_{p^2}$), $\Cm$ and $\Cm^{\ell}$ have the same eigenvectors (up to multiplications by scalars). This is due to the fact that when an eigenvalue is not repeated (multiplicity 1), its eigenvector is unique (up to multiplication by scalars). If $\alpha^{\ell} = \alpha^{p\ell}$, we have that $\Cm^{\ell} = \alpha^{\ell}\Id$ (i.e., scaled identity matrix) with $\alpha^{\ell} \in \FF_{p}$  from (\ref{eq:Cm}). Therefore, we have the following results:
\begin{itemize}
\item  If $\Gamma^{-1}(\Qm) \notin \FF_{p}$ then $\Qm$ has the same eigenvectors of $\Cm$
\item If $\Gamma^{-1}(\Qm) \in \FF_{p}$ then there exist an eigenvector of $\Qm$ that is not an eigenvector of $\Cm$.
\end{itemize} Based on these observations, we prove the theorem statements as follows:

\textbf{Proof of (a):} Since $\Qm$ is a scaled identity matrix, we can choose $\vv_{3}$, not an eigenvector of $\Cm$. Therefore, $\vv_{3}$ is not the eigenvector of $\Qm_{k}$ since $\Gamma^{-1}(\Qm_{k}) \notin \FF_{p}$. This guarantees that the rank conditions hold.

\textbf{Proof of (b):} In this case, any eigenvector of $\Qm$ is also the eigenvector of $\Qm_{k}$ for $k=1,2,3$. Thus,
if we choose $\vv_{3}$ as an eigenvector of $\Qm$ to satisfy the alignment conditions, the rank conditions do not hold, i.e., the desired signal is also aligned with interference signals.
\end{IEEEproof}

In Theorem~\ref{thm3}, case (a) holds for symmetric channels where all interference channels have the same coefficients. However, in general, case (b) occurs with high probability when $\Qm_{\ell k}$ is independently and uniformly drawn from $\Cc^{\star}(p,2)$ and $p$ is large.
Thus, the eigenvector-based IA scheme is generally not feasible.
The following corollary shows that the negative result also holds with high probability for general $m\geq 2$.

\begin{corollary} For the $3$-user scalar IC over $\FF_{p^m}$, if $\deg(\pi_{\beta}(x)) = m$ with $\beta = \Gamma^{-1}(\Qm)$ then the eigenvector-based IA scheme is not feasible.
\end{corollary}
\begin{IEEEproof} Similarly to the case of $m=2$, the companion matrix $\Cm$ is diagonalizable as
\begin{equation}
\Cm = \Am^{-1}\diag(\alpha,\alpha^{p},\ldots,\alpha^{p^{m-1}})\Am.
\end{equation}
Letting $\Qm=\Cm^{\ell}$, we have that $\Qm = \Am^{-1}\diag(\beta,\beta^{p},\ldots,\beta^{ p^{m-1}})\Am$ where $\beta = \alpha^{\ell} = \Gamma^{-1}(\Qm)$. Since $\deg(\pi_{\beta}(x))=m$ by the assumption, the $\beta,\beta^{p},\ldots,\beta^{ p^{m-1}}$ are distinct roots of $\pi_{\beta}(x)$. Thus, $\Qm$ has the same eigenvector with $\Cm$ (up to multiplication by scalars). Also, they are the eigenvector of $\Qm_{k}$ for $k=1,2,3$. Therefore, the rank conditions do not hold and this completes the proof.
\end{IEEEproof}

\subsection{Three Unicast Flows over Wired Networks}

\begin{figure}[ht]
\centerline{\includegraphics[width=10cm]{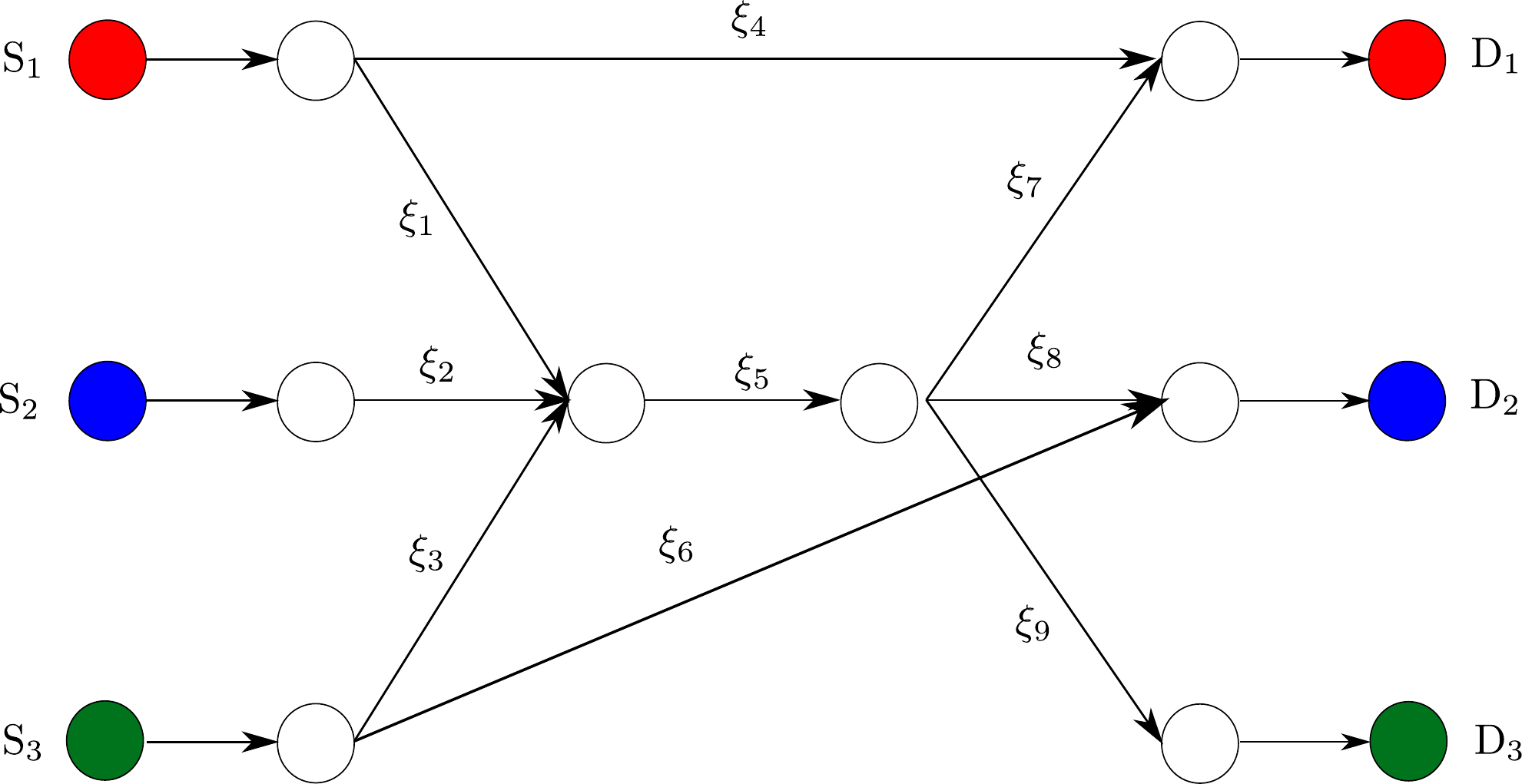}}
\caption{Wired network model as 3-user IC.}
\label{wired-3user}
\end{figure}

For three unicast flows over wired network, a {\em precoding-based network alignment} (PBNA) scheme has been proposed in \cite{Meng,Das}. The main idea of this scheme is that linear network coding at relays can emulate the superpositions in the wireless channel, as shown in Section~\ref{sec:PBNA-2}. Thus, the entire wired network can be seen as a 3-user IC over a finite-field.
This analogy enables us to employ IA schemes, where the only non-trivial processing is performed at the source and destination nodes.
In \cite{Meng,Das}, {\em scalar} linear RLNC was considered, with time-varying coding coefficients $\zeta_{i}[t] \in \FF_{p}$.
This results in a $3$-user IC over $\FF_{p}$ with time-varying channels.
As an example, we consider the network topology as illustrated in Fig.~\ref{wired-3user}.
The channel matrix between source 1 and destination 1 is determined as function of coding variables associated with the paths from source 1 to destination 1:
\begin{equation}
\Hm_{11} = \diag(\xi_{1}[1]\xi_{4}[1]\xi_{6}[1],\ldots,\xi_{1}[\mbox{T}]\xi_{4}[\mbox{T}]\xi_{6}[\mbox{T}])
\end{equation}
where $\xi_{i}[t] \in \FF_{p}$. The other channel matrices can be obtained in the same manner.
For the 3-user IC with $\Hm_{\ell k}$ as channel matrices, the Cadambe and Jafar scheme\footnote{This alignment scheme is different from the eigenvector-based IA scheme and is originally developed for $K$-user {\em scalar} IC with time-varying channel\cite{Cadambe08}.}
in \cite{Cadambe08} can achieve the the sum-rate:
\begin{equation}
R_{\rm{sum}} = \frac{3n+1}{2n+1} \log{p}
\end{equation}
where $\mbox{T} = 2n+1$ is the precoding length (latency).
Namely, this scheme can asymptotically achieve the sum-rate of $\frac{3}{2}\log{p}$ for $\mbox{T} \rightarrow \infty$.
In order to guarantee the feasibility conditions with high probability, $p$ must be sufficiently large \cite{Meng}.
Thus, this scheme may not be applicable when the wired link capacity is not large enough (i.e., $p$ is small).
We will show that this problem can be addressed by using vector linear network coding with coding coefficients from $\Cc^{\star}(p,m)$.
This scheme operates over $m\mbox{T}$ time slots. For every $m$ time slots, the relays perform vector linear network coding by choosing the coding coefficients
$\zeta_{i}[t] \in \Cc^{\star}(p,m)$, for $t=1,\ldots,\mbox{T}$. In our example, the channel matrix becomes the form of $m\mbox{T}\times m\mbox{T}$ block-diagonal matrix:
\begin{equation}
\Hm_{11} = \diag(\xi_{1}[1]\xi_{5}[1]\xi_{7}[1]+\xi_{4}[1],\ldots,\xi_{1}[\mbox{T}]\xi_{5}[\mbox{T}]\xi_{7}[\mbox{T}]+\xi_{4}[\mbox{T}])
\end{equation}
where $\xi_{i}[t] \in \Cc^{\star}(p,m)$.
Using the matrix representation of extension-field in Section~\ref{sec:MAC}, we have:
\begin{eqnarray}
\Gamma^{-1}(\Hm_{11}) &=& \diag(\Gamma^{-1}(\xi_{1}[1]\xi_{5}[1]\xi_{7}[1]+\xi_{4}[1]),\ldots,\Gamma^{-1}(\xi_{1}[\mbox{T}]\xi_{5}[\mbox{T}]\xi_{7}[\mbox{T}]+\xi_{4}[\mbox{T}]))\\
&=& \diag(\Qc_{1}[1]\Qc_{5}[1]\Qc_{7}[1]+\Qc_{4}[1]),\ldots,\Qc_{1}[\mbox{T}]\Qc_{5}[\mbox{T}]\Qc_{7}[\mbox{T}]+\Qc_{4}[\mbox{T}])
\end{eqnarray}
where $\Qc_{i}[t] = \Gamma^{-1}(\xi_{i}[t]) \in \FF_{p^m}^{\star}$.
Then, the end-to-end channel matrix has the form of a $\mbox{T} \times \mbox{T}$ diagonal matrix with diagonal elements from $\FF_{p^m}$
as if the scalar linear network coding is performed over the extension-field $\FF_{p^m}$.
Using this approach, the PBNA scheme can be performed over the larger field $\FF_{p^m}$.
Letting $m$ large, the feasibility conditions of PBNA can be satisfied with high probability regardless of the size
of original field $\FF_{p}$.  Thus, we conclude that the results in \cite{Meng} also hold for any prime $p$.

The use of PBNA simplifies the coding scheme since encoding/decoding operations are determined regardless of the network topology.
Further, this scheme can asymptotically achieve the one half of minimum cut between each source and its intended destination,
which is better than time-sharing.
Yet, this scheme requires long latency (i.e., coding block length) in order to achieve the cut-set bound.
In order to reduce latency, we can employ vector linear network coding with coding coefficients from $\mbox{GL}(p,m)$ (i.e., $\xi_{i} \in \mbox{GL}(p,m)$.
This problem yields a $3$-user IC with channel matrices $\Hm_{\ell k} \in \mbox{GL}(p,m)$.
The coding variables should be chosen carefully, such that the resulting $\Hm_{\ell k}$ are full rank.
Assuming that this holds, the eigenvector-based IA scheme can be used with symbol extension as explained before (see Section~\ref{subsec:3user-MIMO}). In particular, by choosing $m=2$, this scheme achieves $1/2$ of the maximum rate within $4$ time slots since we require at most 2 symbol-extension
(i.e., 2 time slots) for the eigenvalue-based alignment construction.
Hence, the latency of the original PBNA can be greatly reduced.

\section{Concluding Remarks}\label{sec:con}

Linear interference networks over finite fields provide simple yet non-trivial models for 
information theoretic analysis. There are several motivations that make these models also interesting in practice. As reviewed in 
Section \ref{sec:intro}, these models may arise when including the receiver ADC as part of the physical channel, therefore producing an intrinsically 
discrete-input discrete-output channel, or when constraining the relay functions at intermediate nodes to perform Compute and Forward and the
source/destination nodes to use lattice coding/decoding. Also, in wired deterministic networks where the intermediate nodes perform
random linear network coding, the whole network can be compacted as a single linear MIMO transfer function over some finite field. 
In this work we have considered three basic finite field models: $m$-th extension-field scalar models, 
ground-field scalar models with symbol extension, and ground-filed MIMO models. 
We presented a framework to convert a $K$-user linear deterministic multiple access channel (MAC) over $\FF_{p^m}$ with single input/output into
a $K$-user MAC over ground-field $\FF_{p}$ with $m$ multiple inputs/outputs (MIMO).
We showed that the $m\times m$ matrices or the transformed MIMO channel
are represented by the powers of the {\em companion matrix} of a primitive polynomial that generates the extension-field $\FF_{p^m}$. 
Thus, commutativity follows immediately as a consequence of the field multiplicative matrix representation.  
Our framework allows to develop coding schemes for MIMO channels as done in symbol-extension (i.e., coding over multiple time slots) 
for time-varying channels. We focused on the $2\times 2\times 2$ IC topology and on the 3-user IC topology. 
For the $2\times 2\times 2$ scalar IC channel over $\FF_{p^m}$ we showed that the sum rate $(2m-1)\log{p}$ is achievable by 
aligned network diagonalization over the transformed MIMO IC, under certain feasibility condition on the channel coefficients that 
we showed to hold with probability 1 if the channel coefficients are uniformly and independently drawn from the non-zero
elements of $\FF_{p^m}$ and either $m$ or $p$ goes to infinity.  
For the $2\times 2\times 2$ MIMO IC over $\FF_{p}$, we showed that symbol-extension (i.e., coding over multiple time slots) is required in order 
to employ aligned network diagonalization, and we related the dimension of the symbol extension to the extension order of the splitting field
of the characteristic polynomial of a certain system matrix, function of the channel coefficients. 
This observation motivates us to propose a {\em vector} random linear network coding (RLNC) for two-flow multihop wired networks where
the coding matrices are chosen from the powers of a companion matrix of a primitive polynomial that generates the extension-field $\FF_{p^m}$. 
We showed that vector RLNC  achieves the cut-set upper bound for certain network topologies.
In contrast, {\em scalar} RLNC can achieve the same bound only when $p$ is sufficiently large (i.e., the capacity of a wired links is large enough).
This represents a very strong argument in favor of {\em vector} versus {\em scalar} RLNC. 
As an application of finite-field study, we also considered the $2\times 2\times 2$ Gaussian IC with {\em constant} channel coefficients and 
present two coding schemes the combination of which outperforms all other known schemes for any finite SNR. 

For the 3-user MIMO IC over $\FF_{p}$, we show that eigenvector-based IA is also feasible with an appropriate symbol extension.
Yet, we give a negative result for the feasibility of the {\em scalar} 3-user IC over $\FF_{p^m}$. 
showing that eigenvector-based IA  is not feasible although we can always find eigenvalues in an extension-field
and coding/decoding can be performed over symbol-extension by going to the extension-field.
Finally, we show that using vector linear network coding based on the companion matrix approach
for precoding-based network alignment schemes  has some important advantages in terms of feasibility probability 
and  latency.

While this paper represents only a first set of progresses, we believe that the systematic study of linear finite field deterministic networks
is a promising area both in terms of potential theoretical advances and in terms of possible applications.

\section*{Acknowledgment}

This work was supported by NSF Grant CCF 1161801.


\end{document}

%% file: macros.tex
\long\def\comment#1{}

\newfont{\bbb}{msbm10 scaled 700}

\newfont{\bb}{msbm10 scaled 1100}
\newcommand{\CC}{\mbox{\bb C}}
\newcommand{\PP}{\mbox{\bb P}}
\newcommand{\RR}{\mbox{\bb R}}

\newcommand{\ZZ}{\mbox{\bb Z}}
\newcommand{\FF}{\mbox{\bb F}}

\newcommand{\bv}{{\bf b}}
\newcommand{\cv}{{\bf c}}
\newcommand{\dv}{{\bf d}}
\newcommand{\ev}{{\bf e}}

\newcommand{\hv}{{\bf h}}

\newcommand{\tv}{{\bf t}}
\newcommand{\uv}{{\bf u}}
\newcommand{\wv}{{\bf w}}
\newcommand{\vv}{{\bf v}}
\newcommand{\xv}{{\bf x}}
\newcommand{\yv}{{\bf y}}
\newcommand{\zv}{{\bf z}}
\newcommand{\zerov}{{\bf 0}}
\newcommand{\onev}{{\bf 1}}

\newcommand{\Am}{{\bf A}}
\newcommand{\Bm}{{\bf B}}
\newcommand{\Cm}{{\bf C}}

\newcommand{\Em}{{\bf E}}

\newcommand{\Gm}{{\bf G}}
\newcommand{\Hm}{{\bf H}}
\newcommand{\Id}{{\bf I}}
\newcommand{\Jm}{{\bf J}}

\newcommand{\Qm}{{\bf Q}}

\newcommand{\Sm}{{\bf S}}
\newcommand{\Tm}{{\bf T}}

\newcommand{\Vm}{{\bf V}}
\newcommand{\Xm}{{\bf X}}
\newcommand{\Ym}{{\bf Y}}

\newcommand{\Cc}{{\cal C}}
\newcommand{\Dc}{{\cal D}}

\newcommand{\Lc}{{\cal L}}
\newcommand{\Mc}{{\cal M}}
\newcommand{\Nc}{{\cal N}}

\newcommand{\Qc}{{\cal Q}}

\newcommand{\Sc}{{\cal S}}

\newcommand{\Wc}{{\cal W}}
\newcommand{\Vc}{{\cal V}}
\newcommand{\Xc}{{\cal X}}
\newcommand{\Yc}{{\cal Y}}

\newcommand{\lambdav}{\hbox{\boldmath$\lambda$}}

\newcommand{\diag}{{\hbox{diag}}}
\renewcommand{\det}{{\hbox{det}}}
\newcommand{\trace}{{\hbox{tr}}}

\newcommand{\SNR}{{\sf SNR}}

\renewcommand{\Re}{{\rm Re}}
\renewcommand{\Im}{{\rm Im}}
\newcommand{\eqdef}{\stackrel{\Delta}{=}}

\newcommand{\herm}{{\sf H}}

\newcommand{\transp}{{\sf T}}
